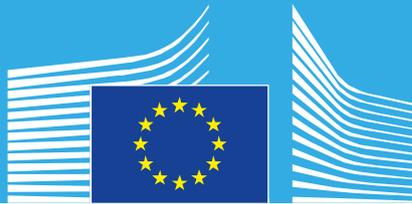

JRC SCIENCE FOR POLICY REPORT

# Subvention des intrants agricoles au Sénégal

*Analyse comparative de trois modes d'interventions à l'aide d'un modèle de ménage agricole*

Ricome Aymeric, Louhichi Kamel, Gomez-y-Paloma Sergio

2020

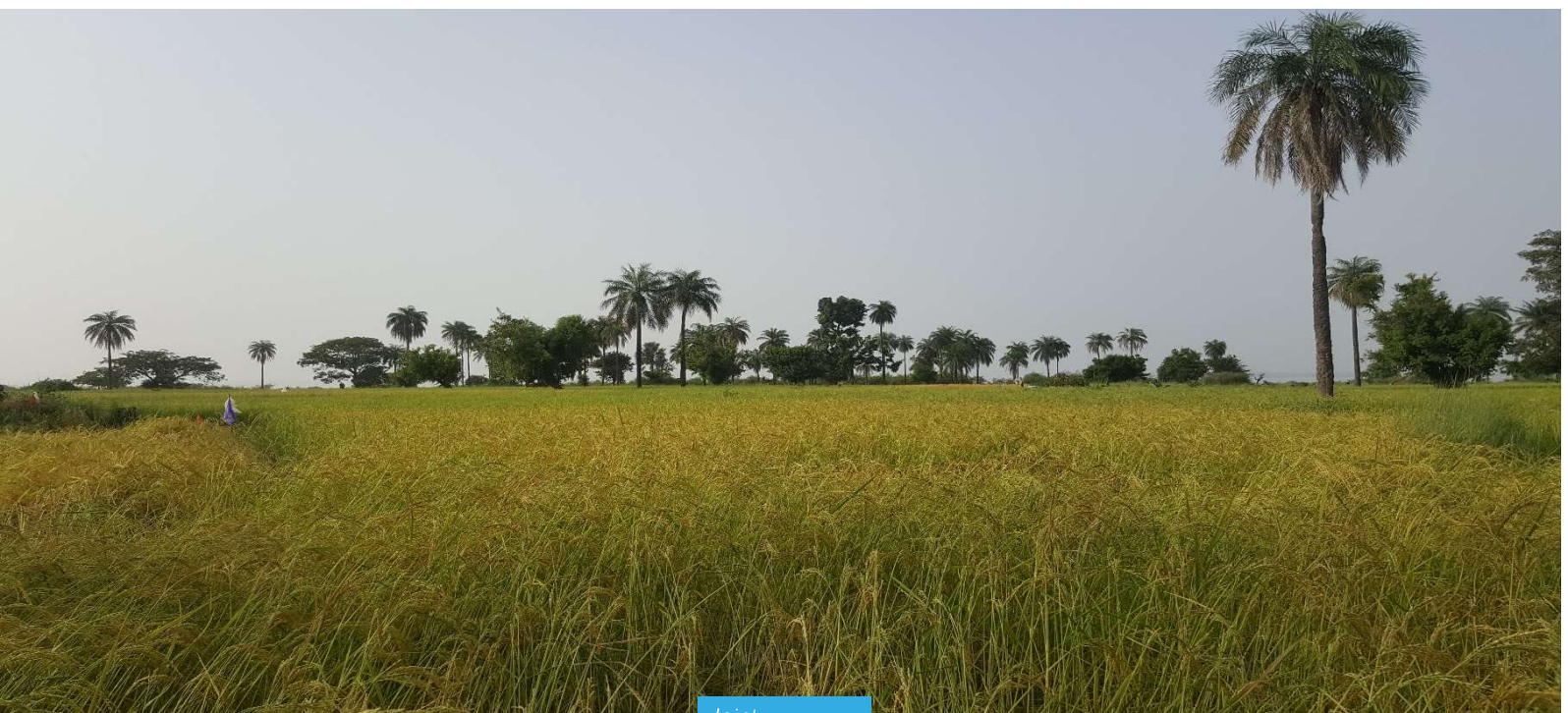

EUR 30238 FR

This publication is a Science for Policy report by the Joint Research Centre (JRC), the European Commission's science and knowledge service. It aims to provide evidence-based scientific support to the European policymaking process. The scientific output expressed does not imply a policy position of the European Commission. Neither the European Commission nor any person acting on behalf of the Commission is responsible for the use that might be made of this publication. For information on the methodology and quality underlying the data used in this publication for which the source is neither Eurostat nor other Commission services, users should contact the referenced source. The designations employed and the presentation of material on the maps do not imply the expression of any opinion whatsoever on the part of the European Union concerning the legal status of any country, territory, city or area or of its authorities, or concerning the delimitation of its frontiers or boundaries.

**Contact information**
Ricome Aymeric
European Commission, Calle Inca Garcilaso 3, 41092 Sevilla, Spain
Email: aymeric.ricome@ec.europa.eu
Tel.: +34 954 48 71 94

**EU Science Hub**
https://ec.europa.eu/jrc

JRC120454

EUR 30238 FR

| | | | |
|---|---|---|---|
| PDF | ISBN 978-92-76-19179-7 | ISSN 1831-9424 | doi:10.2760/511453 |
| Print | ISBN 978-92-76-19178-0 | ISSN 1018-5593 | doi:10.2760/297887 |

Luxembourg: Publications Office of the European Union, 2020

Summary: This report presents the results of an ex-ante impact assessment of several scenarios related to the farmer targeting of the input subsidy programme currently implemented in Senegal. This study has been achieved with the agricultural household model FSSUM-Dev, calibrated on a sample of 2 278 farm households from the ESPS-2 survey. The impacts on crop mix, fertilizer application, farm income and on the government cost are presented and discussed.

# Subvention des intrants agricoles au Senegal

Analyse comparative de trois modes d'interventions à l'aide d'un modèle de ménage agricole

# Table des matières

## Résumé

Depuis maintenant plus de 10 ans, le Sénégal a mis en place un programme de subvention des intrants agricoles qui permet aux agriculteurs bénéficiaires de disposer d'engrais, de semences, et même de matériels agricoles à des prix réduits (pour les semences et l'engrais, cette réduction peut atteindre les 50% du prix de marché). Néanmoins, avec des volumes annuels d'intrants subventionnés relativement limités au regard du nombre de ménages agricoles dans le pays, les bénéficiaires de ce programme restent faibles. De plus, la manière dont les ménages bénéficiaires sont sélectionnés par les commissions locales de cession s'avère peu transparente et de plus en plus critiquée par les différents acteurs du secteur agricole. C'est pourquoi le gouvernement du Sénégal ainsi que les principaux bailleurs de fonds, dont l'Union Européenne, souhaitent réformer ce programme et notamment le mode de ciblage des agriculteurs. En se concentrant uniquement sur les subventions de l'engrais, ce rapport présente les résultats d'une étude visant à évaluer, à l'aide du modèle de ménage agricole FSSIM-Dev, les effets potentiels de ce programme et de le comparer à deux modes de ciblage alternatifs. Trois scénarios ont été ainsi analysés : i) une suppression totale du programme de subvention permettant ainsi d'apprécier ses effets, ii) une universalisation du programme à l'ensemble des ménages sans aucune distinction mais avec une réduction de moitié du quota d'engrais subventionné par ménage, et enfin iii) un ciblage du programme aux seuls ménages ayant une superficie inférieure à 5 ha. Pour cette évaluation, les données de l'Enquête de Suivi de la Pauvreté au Sénégal réalisée en 2011 ont été mobilisées. Cette enquête est réalisée auprès d'un échantillon représentatif de 2278 ménages agricoles répartis sur l'ensemble du territoire sénégalais.

Les résultats des simulations indiquent des effets peu élevés, quels que soient les programmes (scénarios) simulés, sur les décisions de production des agriculteurs (assolement, demande d'engrais et volume produits) et sur leurs revenus (inférieur à +1%) aussi bien au niveau national que régional. Cependant, au niveau individuel, les effets peuvent être plus prononcés pouvant aller, pour le revenu, jusqu'à +30%. En outre, plusieurs ménages (environ 45% dans le cas d'une universalisation du programme) sont contraints par leur trésorerie de continuer à produire sans engrais malgré les subventions. Les cultures qui bénéficieraient le plus du programme de subvention des engrais en terme de volume seraient principalement le maïs et le riz (+3%), ainsi que le niébé (+4%). Par ailleurs, les petites exploitations et les exploitations vivrières semblent les plus dépendantes des subventions et les plus affectées dans le cas d'une suppression du programme actuel. Enfin, le scénario qui cible les exploitations avec moins de 5 ha semble être le plus efficace. Son ratio bénéfice (mesuré en termes de gain de revenu) / coûts (mesurés en termes de montants de subventions allouées) est de 1.2, soit le plus élevé. Néanmoins, comme expliqué en conclusion de ce rapport, ces résultats doivent être interprétés avec prudence et considérés plutôt comme des indications d'ordre de grandeur et non comme une projection certaine en cas d'application stricte d'un de ces programmes dans le futur.

# Préface

Le Centre Commun de Recherche (CCR ou JRC pour *Joint Research Centre* en anglais) est l'une des directions générales de la Commission européenne. Il compte sept instituts de recherche situés dans cinq États membres de l'UE (Belgique, Allemagne, Italie, Pays-Bas et Espagne). Sa mission est de fournir un soutien scientifique et technique à la conception, à l'élaboration, à la mise en œuvre et au suivi des politiques de l'Union Européenne (y compris les mesures de coopération technique internationale) en répondant aux demandes de celles-ci.

Depuis 2014, le CCR est engagé avec la Direction Générale de la coopération internationale et du développement (DG DEVCO) dans un projet intitulé « Soutien technique et scientifique à l'agriculture et à la sécurité alimentaire et nutritionnelle » (TS4FNS) en Afrique subsaharienne. Les principaux objectifs de ce projet sont (i) d'améliorer les systèmes d'informations existants en matière d'agriculture, de nutrition et de sécurité alimentaire, (ii) de réaliser des analyses économiques visant à orienter la prise de décision dans le domaine des politiques agricoles et de coopération, et (iii) de fournir des conseils scientifiques sur des sujets particuliers concernant l'agriculture durable et la sécurité alimentaire et nutritionnelle.

L'une des activités principales du volet économique du projet est l'évaluation des impacts des politiques agricoles et de coopération, à l'échelle micro et macro-économique. Ce travail s'appuie sur le développement de modèles adaptés aux conditions spécifiques de l'économie des pays d'Afrique subsaharienne. Ces modèles permettent par exemple, au niveau micro-économique, d'estimer les effets sur la pauvreté et les inégalités de revenu d'un programme de subvention d'intrants, ou, au niveau macro-économique, d'analyser les impacts sur la production agricole et la balance commerciale, d'une modification des barrières douanières. L'objectif est à la fois de fournir, à la DG DEVCO ainsi qu'aux délégations de l'Union Européenne, des analyses indépendantes sur les impacts des programmes de coopération et de développement de l'UE, mais aussi d'appuyer les autorités locales dans leurs réflexions sur la conception et la mise en œuvre de leurs politiques agricoles, alimentaires et de développement rural.

Ce projet a récemment abouti à la constitution en novembre 2019 du PANAP (Pan-African Network for economic Analysis of Policies) qui est un réseau soutenu par la DG JRC de la Commission Européenne regroupant des chercheurs de la DG JRC mais aussi des chercheurs et académiques des pays partenaires d'Afrique subsaharienne dont l'objectif est de favoriser des recherches sur l'économie et les politiques agricoles de ces pays.

Enfin, notons que cette présente étude sur l'évaluation ex-ante de l'impact de différents critères de ciblage des exploitations agricole pour le programme de subventions des intrants sera complétée par une évaluation ex-post qui débutera au second semestre 2020.

***Auteurs***

Aymeric Ricome[a]

Kamel Louhichi[a,b]

Sergio Gomez-y-Paloma[a]

[a] Commission Européenne, Centre Commun de recherche (CCR), Séville, Espagne.

[b] INRAE–UMR Economie Publique, Thiverval-Grignon, France

# 1 Introduction

Avec un PIB par habitant de 1 546 USD, le Sénégal est la 4ème économie de la sous-région ouest-africaine. Le pays a connu une forte croissance économique depuis 20 ans puisque son PIB par habitant a progressé de 39% depuis 2000 (USD constant). Par ailleurs, selon le rapport sur le développement humain 2019 établi par le PNUD, l'indice de développement humain (IDH) du Sénégal qui était de 0,39 en 2000 est passé à 0,514 en 2018. Si des progrès sociaux et humains ont été réalisés, le pays reste néanmoins classé au 166$^{ième}$ rang sur 189, toujours selon le PNUD. Il se situe légèrement au-dessus de la moyenne des pays à faible développement humain (0.507) mais se situe en dessous de la moyenne des pays de l'Afrique subsaharienne (0,541). De plus, malgré ces progrès, l'économie sénégalaise reste fragile et la sécurité alimentaire, notamment pour les 60% des 15 millions de sénégalais vivant en milieu rural, reste une préoccupation majeure pour le pays. Le pourcentage de personnes vivant sous le seuil de pauvreté national atteignait 46% en 2011 (Banque Mondiale 2020). La prévalence de la sous-alimentation (pourcentage de la population dont l'apport alimentaire est insuffisant pour satisfaire les besoins énergétiques) dans le pays en 2017 est de 11% bien qu'il ait drastiquement baissé depuis 2000 (28%) (FAO 2019). Le Sénégal demeure un pays agricole dans la mesure où ce secteur emploie environ 46% de la population active et contribue à hauteur de 16% du PIB (Banque Mondiale 2020).

Depuis son accession à l'indépendance, le Sénégal a défini et mis en œuvre successivement plusieurs stratégies de développement agricole afin de faire jouer à ce secteur un rôle moteur dans la croissance économique et pour améliorer la sécurité alimentaire du pays. En dépit des ambitions et des objectifs poursuivis à travers ces politiques, le secteur agricole éprouve toujours du mal aujourd'hui encore à jouer un rôle moteur dans l'économie nationale. Il doit faire face à une série de défis parmi lesquels on trouve le problème de la dégradation des sols dans plusieurs régions du pays, la forte dépendance des systèmes de production aux aléas pluviométriques, la forte croissance démographie, la vétusté et l'insuffisance de matériel agricole, ainsi que la faible disponibilité des engrais et des semences en quantité et en qualité. C'est pour surmonter ces défis mais aussi en réponse à la crise alimentaire mondiale de 2007/08 que l'état sénégalais a fortement renforcé le dispositif de soutien à son secteur agricole en lançant dès 2008 la Grande Offensive Agricole pour la Nourriture et l'Abondance (GOANA), suivi en 2013 par le lancement du Programme d'Accélération de la Cadence de l'Agriculture Sénégalaise (PRACAS). Le budget total que l'Etat Sénégalais a consacré à l'agriculture est ainsi passé de 136 milliards de F CFA en 2010 à 264 milliards en 2014 (IPAR 2015), soit un quasi-doublement du budget agricole en seulement 4 années. Le programme phare du PRACAS pour relancer la production agricole et accompagner les ménages agricoles est le programme de subventions d'intrants agricoles, qui comprend des semences, des engrais et plus récemment du matériel agricole.

Dans le cadre de ce programme, l'Etat subventionne chaque année des intrants, sous contrainte de l'enveloppe budgétaire notifiée par le Ministère en charge des finances, des quantités d'intrants effectivement cédés aux producteurs par les agro-fournisseurs. Des "commissions de cession" à l'échelle des communautés rurales sont en charge de la distribution auprès des agriculteurs de ces intrants subventionnés. Du fait de quantités d'intrants subventionnés relativement limitées, tous les agriculteurs n'en bénéficient pas de sorte qu'un ciblage se réalise de facto au travers des décisions de ces commissions de cession. De plus, le ciblage réalisé n'est pas systématisé et codifié ce qui conduit à une certaine opacité des critères utilisés par ces commissions pour décider du choix des agriculteurs bénéficiaires (IPAR 2015, Feed the Future 2018). Ce manque supposé de transparence dans le ciblage a conduit de nombreux acteurs de la filière à émettre des doutes sur son bien-fondé et au final sur l'efficacité globale du programme de subvention des intrants dans le pays (Seck 2017, Feed the Future 2018).

L'objectif de ce rapport est de présenter les résultats d'une étude portant sur l'évaluation ex ante de l'effet de différents types de ciblage du programme de subvention des engrais sur les ménages agricoles. Les effets sont mesurés en termes d'assolement, d'utilisation d'engrais, et de revenu agricole. Les coûts budgétaires sont également mesurés. Pour réaliser cette évaluation, le modèle de ménage agricole FSSIM-Dev (Farming System Simulator for Developing Countries) a été utilisé en s'appuyant sur les données issues de la seconde Enquête de Suivi de la Pauvreté au Sénégal 2 (ESPS2). Il s'agit d'une enquête réalisée en 2011 par l'Agence Nationale de la Statistique et de la Démographie (ANSD) dans l'ensemble du pays auprès de 2 278 ménages agricoles.

Ce rapport est organisé comme suit : la seconde section présente un bref aperçu du secteur agricole ainsi que les principaux programmes de subventions aux intrants au Sénégal. La troisième section décrit le modèle de ménage FSSIM-Dev notamment sa méthodologie, ses spécificités, sa formulation mathématique et la méthode de calibrage employée. La section 4 détaille les données de l'enquête utilisées dans le modèle FSSIM-Dev ainsi que les scénarios retenus. La cinquième section expose les résultats de la simulation des impacts des programmes de subventions aux engrais, et la dernière section conclut.

## 2 Le secteur agricole du Sénégal et la politique de subvention des intrants

### 2.1 Un rapide aperçu des systèmes de production agricole

Les exploitations agricoles au Sénégal sont principalement de types familiales, c'est à dire que la main d'œuvre utilisée provient essentiellement du ménage et les produits de la récolte sont principalement destinés à la consommation alimentaire des membres du ménage. On dénombrait en 2013 autour de 755 350 exploitations de ce type, réparties sur tout le territoire (Sall 2015). A côté de ces exploitations familiales, coexistent aussi des exploitations dites *commerciales*, généralement plus grande que les exploitations familiales, mobilisant au contraire des travailleurs salariés et dont la logique de production est essentiellement marchande. Elles restent minoritaires (occupent moins de 5% du territoire) de sorte que ce sont les exploitations familiales qui contribuent à l'essentiel de la production nationale.

Sur les 19.7 millions d'hectares (ha) que couvre le pays, la superficie considérée cultivable est estimée à 3.8 millions d'ha, dont 2.4 millions sont effectivement cultivés chaque année (Rioux et al. 2011). La pluviométrie du pays est faible pour la grande majorité du territoire malgré l'existence d'un gradient de pluviométrie nord-sud. C'est ainsi que sur la période 1991-2010, l'extrême nord du pays a reçu en moyenne moins de 300 mm de pluie alors que l'extrême sud (la Casamance) a bénéficié de beaucoup plus de pluies, allant jusqu'à 1200 mm par an. En dépit d'une forte contrainte pluviométrique, moins de 5% de terres sont irrigables dans le pays (FAO, 2019). L'agriculture sénégalaise repose essentiellement sur les précipitations qui débutent au mois de juin/juillet et se terminent en septembre/octobre.

Les systèmes de production au Sénégal sont très hétérogènes et dépendent des zones agro-écologiques dans lesquels ils se trouvent. On distingue en général 6 grandes zones agro-écologiques dans le pays avec des potentiels agricoles variés en rapport principalement avec le climat, le sol, la végétation :

— La vallée du fleuve Sénégal avec une forte présence de cultures irriguées (riz et horticulture) et des cultures de décrue (maïs, sorgho)

— La zone sylvo-pastorale (Ferlo) principalement dédiée aux activités pastorales

— Le bassin arachidier considéré comme le grenier du Sénégal avec la production de céréales sèches (mil, sorgho) et de légumineuses (arachide, niébé)

— Les niayes, zone côtière avec prépondérance de productions horticoles

— Le Sénégal oriental, zone de polyculture (mil, sorgho, maïs, coton, arachide…)

— La casamance, zone de culture de riz pluvial mais aussi de céréales sèches (mil, sorgho, maïs…) et de cultures de rente (arachide, arboriculture fruitière)

L'agriculture sénégalaise dispose d'un important potentiel pour améliorer la sécurité alimentaire du pays, mais sa production reste encore très dépendante des aléas climatiques (pluies) et économiques (volatilité des prix des produits agricoles, débouchés incertains). Elle montre néanmoins une forte capacité de résilience, grâce notamment à la diversification de ces activités agricoles.

### 2.2 L'agriculture au cœur de la politique gouvernementale

Depuis son adoption formelle en novembre 2012, le Plan Sénégal Emergent (PSE) constitue le cadre de référence des politiques de développement économique et sociale sur le moyen (2023) et le long terme (2035) au Sénégal. Dans ce document, la modernisation de l'agriculture est considérée comme une priorité pour assurer une transformation de l'ensemble de l'économie du pays. C'est ainsi que l'essentiel de la stratégie du gouvernement en matière agricole repose sur le Programme d'Accélération de la Cadence de l'Agriculture Sénégalaise (PRACAS), le volet agricole du PSE. Le PRACAS vise à assurer le développement de l'agriculture, de l'élevage, de la pêche, de l'aquaculture et de l'industrie agroalimentaire, le renforcement de la sécurité alimentaire du pays et le rééquilibrage de la balance commerciale, le développement de filières intégrées à haute valeur ajoutée et compétitives, et la préservation des équilibres socio-économiques dans les régions rurales (Ministère de l'Agriculture et de l'Equipement Rural 2014). Cette volonté affichée d'appuyer le secteur agricole s'est concrètement traduite par une hausse de la part des dépenses publiques destinées à l'agriculture et à l'alimentation. Ces dépenses (au sens large incluant les soutiens directs et indirects) ont augmenté de 295 milliards de FCFA en 2010 à 495 milliards de FCFA en 2015, soit une hausse de 67%, alors que les dépenses publiques globales n'ont augmenté sur le même temps que de 30% (Boulanger et al. 2018). Le détail des dépenses publiques dans ce secteur est présenté dans la Figure 1.

**Figure 1**: Dépenses publiques (ressources propres et externes) décaissées en soutien à l'agriculture et l'alimentation entre 2010 et 2014 (milliards de FCFA)

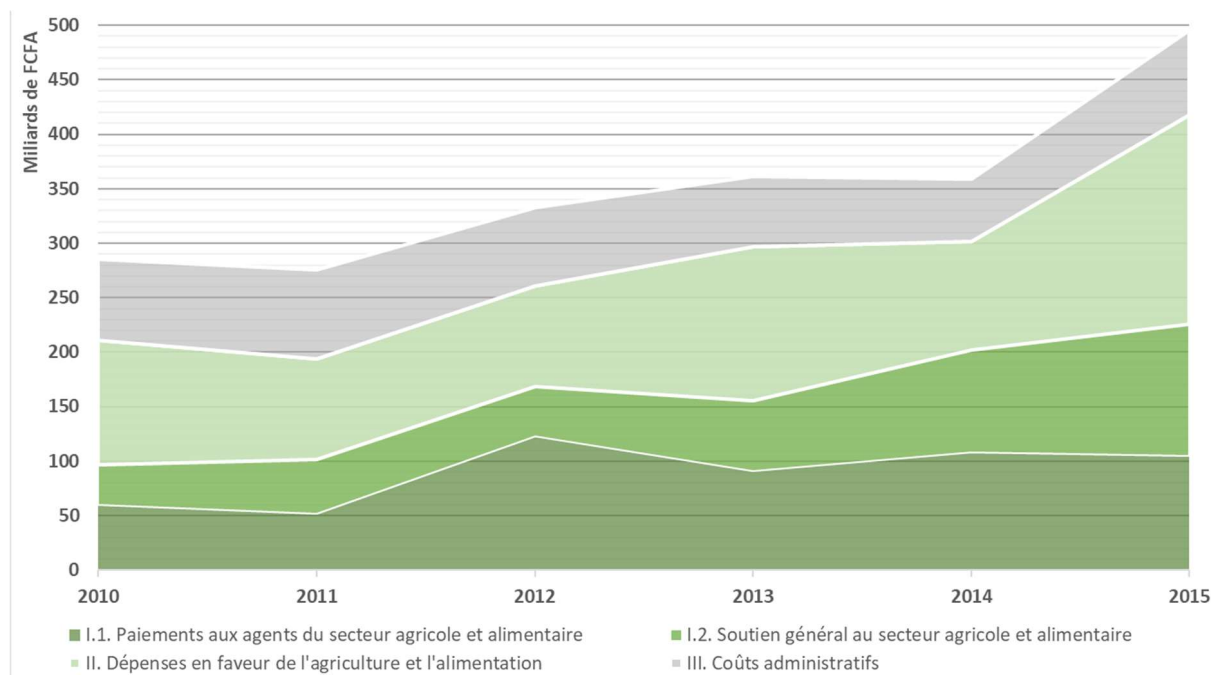

Source : Boulanger et al., 2018

Parmi les objectifs chiffrés avancés par le gouvernement dans le cadre du PRACAS figure celui d'atteindre une production de 1 600 000 tonnes de paddy, de 350 000 tonnes d'oignon et enfin de 1 000 000 tonnes d'arachide. Pour atteindre ces objectifs, le PRACAS propose une stratégie d'ensemble et des stratégies spécifiques aux filières. La stratégie d'ensemble se décline en cinq axes d'intervention (Ministère de l'Agriculture et de l'Equipement Rural 2014) :

— L'intensification des productions agricoles

— La maîtrise de l'eau

— La recherche agricoles – formation – conseil agricole et rural

— La valorisation et la mise en marché des produits agricoles

— La gestion de la qualité (conformité des produits sénégalais aux normes commerciales, sanitaires et phytosanitaires exigées par les marchés de destination)

C'est sous le premier axe que se concentre l'essentiel des transferts directs aux producteurs avec pour politique principale la mise en place d'un programme de subvention des intrants.

## 2.3   Le programme de subvention des intrants agricoles

Selon le document officiel du PRACAS, l'objectif de ce programme est de cibler toutes les filières agricoles et doit permettre la subvention partielle de 50 000 à 80 000 tonnes d'engrais toutes formules confondues. Le programme repose également sur l'amélioration de l'accès des producteurs aux intrants en encourageant la mise en place de points de vente en période de commercialisation des produits, notamment de l'arachide (Ministère de l'Agriculture et de l'Equipement Rural 2014). Par ailleurs, ce programme de subvention au Sénégal existe en réalité depuis la campagne agricole 2004/05 mais il ne s'est fortement renforcé qu'en 2012/2013 en augmentant très significativement les quantités de semences et d'engrais subventionnées ainsi que leurs taux de subvention (autour de 50% même si ce taux peux changer à chaque campagne agricole) mais aussi en élargissant ce programme de subvention au matériel agricole. C'est ainsi que l'enveloppe budgétaire allouée à ce programme n'a cessé d'augmenter (Tableau 1), passant pour les seuls engrais de 4.6 milliards de FCFA en 2004/05 à 18 milliards de FCFA en 2012/13 (IPAR, 2015).

**Tableau 1**: évolution de la dépense totale allouée à la subvention des engrais entre 2004 et 2013 (milliards FCFA)

| 2004/05 | 2005/06 | 2006/07 | 2007/08 | 2008/09 | 2009/10 | 2010/11 | 2011/12 | 2012/13 | 2013/14 |
|---------|---------|---------|---------|---------|---------|---------|---------|---------|---------|
| 4.6 | 4.2 | 4.3 | 7.8 | 9.1 | 7.7 | 9 | n/a | 18 | 13.9 |

Source: IPAR, 2015

Dans les détails, le programme de subvention sénégalais fonctionne de la manière suivante. Chaque année, l'Etat évalue les besoins " théoriques " en engrais pour chaque culture selon les objectifs de production régionaux, les emblavures prévues et les doses d'engrais recommandées pour atteindre les objectifs de production par culture (Tableau 2). Ensuite, l'Etat subventionne, sous contrainte de l'enveloppe budgétaire notifiée par le Ministère en charge des finances, les quantités d'intrants effectivement cédés aux producteurs par les fournisseurs d'intrants. La distribution reste néanmoins à la charge des fournisseurs. Les semences certifiées subventionnées sont quant à elles distribuées par des semenciers privés ou directement par l'ISRA, l'organisme public de recherche agricole, qui a en charge la production des pré-bases de qualité. Une fois que les quantités d'engrais et de semences subventionnées sont connues au niveau régional, celles-ci sont réparties au niveau de chaque communauté rurale où une "commission de cession" se charge de les répartir entre les agriculteurs. Chaque agriculteur bénéficiant du programme de soutien peut recevoir jusqu'à 3 sacs de 50 kg d'engrais subventionné. Les niveaux exacts de subvention varient selon le produit (engrais, semences, matériel agricole), la culture visée et enfin selon les années. Par exemple pour la campagne 2013/14, le taux de subvention moyen des semences était de 76% du prix du marché. Pour la campagne 2014/15, ce taux moyen est passé à 67% (IPAR, 2015).

**Tableau 2**: Formules d'engrais par spéculation et doses préconisées par l'ISRA

| Culture | Formule | Dose préconisée par l'ISRA (kg/ha) |
|---------|---------|------------------------------------|
| Riz | NPK 18.46.0 | 100 |
|  | Urée | 350 |
| Arachide | NPK 6.20.10 | 150 |
| Mil/Sorgho | NPK 15.10.10 | 150 |
|  | Urée | 100 |
| Maïs | NPK 15.15.15 | 200 |
|  | Urée | 200 |
| Horticulture | NPK 9.23.30 | 400 |
|  | NPK 10.10.20 | 200 |
|  | Urée | 50 |

Source: République du Sénégal, 2014

Cependant, du fait de quantités d'intrants subventionnés relativement limitées, tous les agriculteurs ne bénéficient pas de subvention et la sélection entre agriculteurs est établie par les commissions de cession. Les critères utilisés par ces commissions sont peu connus et probablement variables selon les commissions étant donné qu'elles sont constituées de responsables politiques locaux, des autorités administratives, des responsables d'organisations villageoises, etc. (IPAR, 2015)

Ce manque supposé de transparence dans le ciblage a conduit de nombreux acteurs de la filière à émettre des doutes sur son bien-fondé et au final sur l'efficacité globale du programme de subvention aux intrants dans le pays. Le gouvernement sénégalais ainsi que les bailleurs de fond participant au financement de ce programme souhaitent réformer ce programme et notamment le mode de ciblage des agriculteurs afin de le rendre plus transparent et plus équitable.

Ce rapport présente les résultats d'une étude visant à évaluer, à l'aide du modèle de ménage agricole FSSIM-Dev, les effets potentiels du programme de subvention actuellement en place au Sénégal et de le comparer avec deux types de ciblage alternatifs. Néanmoins, l'étude se bornera ici au engrais et ne tiendra pas compte des autres intrants eux-aussi subventionnés (semences et matériel agricole).

# 3   Le modèle FSSIM-Dev

L'analyse de l'impact des différents types de ciblage du programme de subvention repose sur un modèle de simulation du comportement des ménages agricoles qui permet de quantifier les effets en termes d'assolement, de production agricole, de revenu agricole, et donc *in fine*, de réduction de la pauvreté en milieu rural. Le modèle permet aussi d'estimer le coût budgétaire de chaque type de ciblage simulé.

Les modèles de ménages agricoles sont très souvent utilisés pour réaliser de telles analyses, en particulier dans les pays en voie de développement car ils permettent de considérer la non-séparabilité des décisions de production et de consommation des ménages agricoles, l'hétérogénéité de ces mêmes ménages, la saisonnalité de la production agricole et permet enfin de considérer les pratiques agricoles utilisées par les agriculteurs de manière explicite (Tillie et al. 2018). Par ailleurs, l'intérêt d'un modèle de ménage agricole est qu'il permet d'estimer l'impact d'une mesure sur toutes les exploitations d'un échantillon, et donc il permet à l'analyse de prendre en compte la façon dont les effets se distribuent dans toutes les exploitations, et pas seulement leurs effets moyens. Louhichi et Gomez-y-Paloma (2014) ont récemment passé en revue toutes les études basées sur un modèle des ménages agricoles dans les pays en développement, soulignant les avantages et les inconvénients des différentes méthodologies, couvertures géographiques et hypothèses comportementales utilisées.

Le modèle utilisé ici est le *Farming System SIMulator for Developing countries* (FSSIM-Dev) qui est particulièrement adapté aux spécificités des ménages agricoles des pays en développement (Louhichi et al. 2020). Ce modèle micro-économique requiert néanmoins une quantité importante de données individuelles sur les ménages et leurs exploitations agricoles à savoir: l'occupation des sols, la production et les rendements des cultures, la quantité et les prix des intrants, l'utilisation de la main d'œuvre, le prix de vente des produits agricoles, etc.

Avant de présenter les données utilisées, les principales caractéristiques du modèle FSSIM-Dev sont présentées dans cette section. Le lecteur intéressé par plus de détails sur sa structure et son mode de fonctionnement peut trouver toutes ces informations détaillées dans Louhichi et al. (2020).

## 3.1   Aperçu général du modèle FSSIM-Dev

FSSIM-Dev est un outil d'aide à la décision économique destiné à être utilisé dans le contexte des pays en développement pour améliorer les connaissances sur la sécurité alimentaire et le niveau de la pauvreté en milieu rural. Il vise à informer les décideurs politiques et les partenaires de développement sur la façon dont les changements des prix, de technologie, des politiques agricoles et alimentaires pourraient affecter la viabilité et la sécurité alimentaire des ménages agricoles.

FSSIM-Dev est un modèle de ménage conçu pour l'analyse de l'agriculture familiale ou paysanne où les décisions de production, de consommation et d'allocation de la main-d'œuvre sont indissociables en raison des imperfections du marché. Il permet une représentation micro-économique du ménage agricole et, par conséquent, une analyse fine des effets de chocs exogènes sur la viabilité des ménages agricoles situés dans différentes régions du pays. Il s'agit d'une extension du modèle d'exploitation FSSIM développé dans le cadre du projet européen SEAMLESS (van Ittersum et al. 2008) pour analyser l'impact des politiques agricoles et environnementales sur la durabilité des systèmes de production en Europe (Janssen et al. 2010, Louhichi et al. 2010).

Le principal atout du modèle FSSIM-Dev est sa capacité à prendre en compte les principales caractéristiques de l'agriculture des pays en développement à savoir: (i) la non-séparabilité des décisions de production et de consommation, (ii) l'interaction entre les ménages agricoles pour l'utilisation des facteurs de production, (iii) l'hétérogénéité des ménages agricoles par rapport aux paniers de consommation et aux dotations en ressources, (iv) l'interdépendance entre les coûts de transaction et les décisions de participation au marché, et (v) la saisonnalité des activités agricoles et de l'utilisation des ressources.

FSSIM-Dev fonctionne avec des prix exogènes pour représenter les offres des principales activités végétales et animales au niveau des ménages agricoles. Il simule comment un scénario donné, par exemple une nouvelle politique agricole, peut affecter un ensemble d'indicateurs à savoir : l'assolement, l'utilisation des ressources et des intrants, la production végétale et animale, la consommation, les revenus agricoles et des ménages, la sécurité alimentaire et nutritionnelle des ménages, les dépenses publiques et les externalités environnementales telles que l'érosion des sols et/ou les émissions de gaz à effet de serre. Ces indicateurs peuvent être agrégés et comparés selon le type de ménages, l'orientation technique de l'exploitation, la dimension économique ou la région/village d'appartenance. Ces résultats ne peuvent, néanmoins, être considérés comme des projections ou des prévisions dont la réalisation serait certaine mais plutôt comme des indications de tendances suscitées par les chocs exogènes.

La principale motivation pour le développement de ce type d'outil microéconomique est la forte hétérogénéité des politiques agricoles aussi bien en termes de mise en œuvre (c'est-à-dire les politiques sont de plus en plus ciblées et spécifiques à certaines catégories de la population agricole) que d'impacts. En effet, les réponses des agriculteurs aux changements de politiques varient d'un ménage à un autre selon la location, la dotation en ressources, l'utilisation des terres, l'accès aux marchés, le régime foncier, l'âge, le sexe, la situation économique, la composition de la famille, etc. Cela pourrait notamment être le cas, par exemple, lorsqu'il s'agit d'instruments politiques qui provoquent des changements dans la production et la consommation du ménage. L'ampleur et la direction de ces effets dépendront du comportement de chaque agent qui dépend à son tour de ses caractéristiques, ses préférences, sa localisation, etc. Pour prendre en compte l'hétérogénéité entre les exploitations et identifier les gagnants et les perdants des politiques existantes ou alternatives, une analyse au niveau micro est donc nécessaire.

FSSIM-Dev est configuré de façon générique et modulaire pour être facilement adaptable et réutilisable à de nouvelles questions politiques et/ou à différentes conditions biophysiques et socio-économiques. Grâce à son caractère générique et modulaire, le modèle peut être appliqué à des ménages individuels (c'est-à-dire réels) ou représentatifs (c'est-à-dire des ménages typiques ou moyens). Il peut également être utilisé pour l'analyse des décisions des agriculteurs qui sont entièrement orientés vers le marché comme ceux de (semi)subsistance ou qui opèrent dans des marchés imparfaits.

Une première version du modèle a été utilisée en 2014 pour évaluer les effets d'une politique de soutien aux semences de riz sur la viabilité des ménages agricoles en Sierra Leone (Louhichi and Gomez y Paloma 2014) et une seconde version plus avancée a été utilisée pour réaliser des études d'impacts en Côte d'Ivoire (Tillie et al. 2018), en Ethiopie (Louhichi et al. 2019), au Niger (Tillie et al. 2019), et enfin en Tanzanie (Ricome et al. 2020).

## 3.2 Structure et formulation mathématique du modèle FSSIM-Dev

FSSIM-Dev est un modèle d'optimisation statique non-linéaire qui repose à la fois sur le cadre général d'utilité du ménage et sur les contraintes techniques de production agricole, dans un régime non-séparable. Basé sur la programmation mathématique positive (Howitt 1995), FSSIM-Dev maximise une fonction objectif soumise à un ensemble de contraintes de dotation en ressources, de consommation humaine et de politiques agricoles. Il suppose que le ménage agricole maximise son revenu espéré, défini comme le revenu obtenu de toutes les activités économiques d'une famille vivant dans le même ménage à savoir : le revenu agricole, les revenus des facteurs de production commercialisés (salaires non agricoles, loyer des terres et/ou du matériel, etc.) et les revenus extra-agricoles (c'est-à-dire hors exploitation). Le revenu agricole est défini comme la somme des revenus économiques que les ménages agricoles ont obtenus en vendant ou en consommant leurs propres produits agricoles. Les revenus hors-exploitation sont définis d'une manière exogène et peuvent provenir de différentes sources telles que les salaires hors-ferme, le petit commerce, les activités indépendantes, les pensions, les transferts et les dons.

Le revenu agricole est défini comme la somme des marges brutes espérées moins une fonction de comportement non-linéaire (quadratique) propre à chaque activité du ménage agricole. La marge brute est le total des recettes, y compris les ventes et l'autoconsommation, plus les subventions moins les charges opérationnelles. Les charges opérationnelles incluent les coûts de semence, des engrais, des produits phytosanitaires et d'autres coûts spécifiques. La fonction quadratique est une fonction de comportement introduite pour calibrer le modèle à une situation observée, comme c'est généralement le cas dans les modèles de programmation mathématique positive. Cette fonction vise à répliquer d'une manière précise les décisions de production et de consommation des ménages agricoles en captant les effets de facteurs qui ne sont pas explicitement introduits dans le modèle, tels que les coûts du capital, l'aversion au risque, l'anticipation des prix, les erreurs de spécification du modèle, etc. (Paris and Howitt 1998, Heckelei 2002, Henry de Frahan et al. 2007).

La formulation mathématique générale du modèle est la suivante :

(E1) $$Max\ Z_h = \sum_i (s_{h,i} + cs_{h,i})p_{h,i} + \sum_i sb_{h,i}x_{h,i} - \sum_{i,k} a_{h,i,k}x_{h,i} - \sum_i (d_{h,i} + 0.5 Q_{h,i,i'}x_{h,i'})x_{h,i} + \sum_{tf}(s_{h,tf} - b_{h,tf})p_{h,tf} + ExInc_h$$

s.t.

(E2) $$\sum_m A_{i,m}x_i \leq B_m + b_m - s_m \quad [\rho_m]$$

(E3) $$c_{h,j}p_{h,j} = \beta_{h,j}(Z_h - \sum_{j'} \gamma_{h,j'}p_{h,j'}) + \gamma_{h,j}p_{h,j}$$

(E4a) $$p_j^m t_{h,j}^s \leq p_{h,j} \leq p_j^m t_{h,j}^b$$

(E4b) $$p_{tf}^m t_{h,tf}^s \leq p_{h,tf} \leq p_{tf}^m t_{h,tf}^b$$

(E5) $$s_{h,j} b_{h,j} = 0$$

(E6a) $$s_{h,j}(p_{h,j} - p_j^m t_{h,j}^s) = 0$$

(E6b) $$b_{h,j}(p_{h,j} - p_j^m t_{h,j}^b) = 0$$

(E7) $$q_{h,j} + b_{h,j} = s_{h,j} + c_{h,j}$$

(E8) $$c_{h,j} = cs_{h,j} + b_{h,j}$$

où **Z** est le revenu du ménage agricole **h**, **p** est le vecteur (n×1) des prix des biens j et des facteurs de production potentiellement échangeables **tf** (terre, travail) du ménage agricole **h**, **s** est le vecteur (n×1) des quantités de biens vendus ou des facteurs de production cédés, **cs** est le vecteur (n×1) de quantités de biens autoconsommés, **x** est le vecteur (n×1) des activités agricoles optimales **i**, **sb** est le vecteur (n×1) des subventions à la production (le cas échéant), **a** est la matrice (n×k) des charges variables des intrants **k**, **q** est le vecteur (n×1) de quantités de bien produits sur l'exploitation, **b** est le vecteur (n×1) de quantités de biens achetés ou des facteurs de production loués et **c** est le vecteur (n×1) de quantités de biens consommés. γ est le vecteur (n×1) de la consommation incompressible du ménage, **β** est le vecteur (n×1) des préférences du ménage pour le produit **j** et leur somme doit être égale à l'unité, **pᵐ** est le vecteur (n×1) des prix du marché des biens **j** et **tᵇ** et **tˢ** sont respectivement les vecteur (n×1) des coûts de transaction supportés par le ménage lors de l'achat ou de la vente des biens **j** ou des facteurs **tf**. **d** est le vecteur (n×1) de la partie linéaire de la fonction de comportement et **Q** est la matrice (n×n) symétrique et (semi-définie) positive de la fonction de comportement. **A** est la matrice (n×m) de coefficients techniques, **B** est le vecteur (m×1) des dotations initiales en ressources (terre, travail) et ρ est le vecteur (m×1) de leurs valeurs marginales respectives. **Exinc** est un paramètre qui représente le revenu extra-agricole, **w** est le coefficient de pondération du ménage (c'est-à-dire le poids du ménage dans la région), **M** et **R** sont respectivement les quantités de facteur de production et de produits importés et exportés vers/à d'autres régions. **Q**, **d** et ρ sont estimés en utilisant une variante de la programmation mathématique positive (Louhichi et al. 2018). **β** et γ sont estimés en utilisant une approche bayésienne, appelée Highest Posterior Density (Heckelei et al. 2008).

Un certain nombre de contraintes sont prises en compte dans FSSIM-Dev pour (i) modéliser la dotation en ressources de l'exploitation (Eq. 2) (ii) la fonction LES, Linear Expenditure System, ou système de dépenses linéaires représentant la consommation du ménage (E3), (iii) la discontinuité dans la décision de participation aux marchés due à l'existence des coûts de transaction (E4a et E4b) (c'est-à-dire ces coûts de transaction amplifient les prix effectivement payés par les acheteurs et diminuent les prix effectivement reçus par les vendeurs), (iv) les conditions dites de relaxation complémentaire, pour s'assurer, d'une part, que pour chaque produit, un ménage agricole peut être acheteur ou vendeur mais pas simultanément (E5) et, d'autre part, qu'un ménage agricole peut vivre en autarcie et utilise son propre prix (E6a et E6b), et (v) finalement les deux conditions d'équilibre du marché: la première (eq. E7 et E8) garantit l'équilibre des produits au niveau de chaque ménage, c'est-à-dire pour chaque produit la somme de la production et des achats doit être égale à la vente plus la consommation (eq. E7); et la seconde (eq. E8) garantit que le niveau total consommé égal l'autoconsommation et les achats.

**Figure 2.** Structure simplifié du modèle FSSIM-Dev

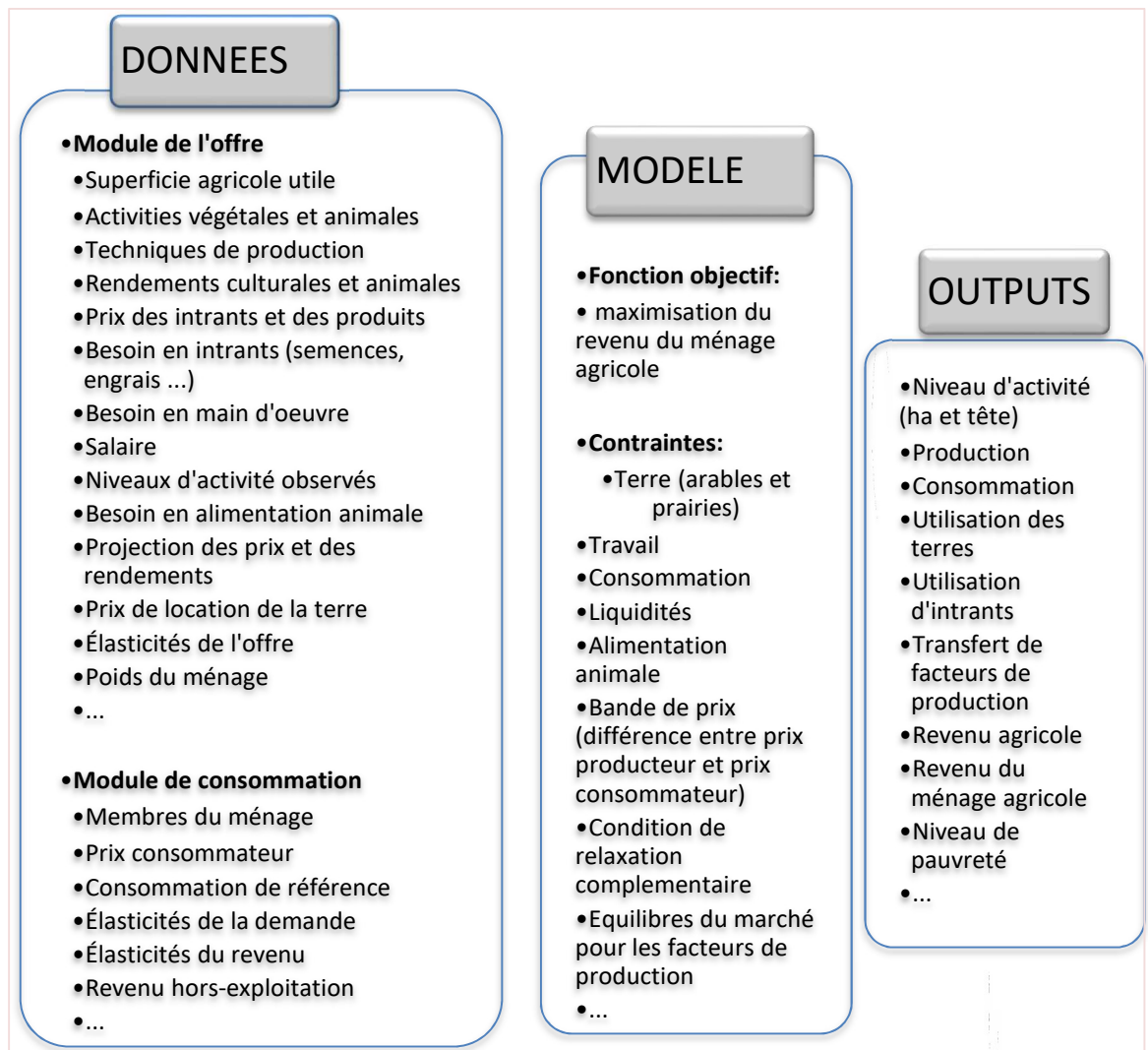

Source: les auteurs.

Une maquette unique a été appliquée à tous les ménages agricoles modélisés afin d'assurer une gestion uniforme des modèles individuels et de leurs résultats. Autrement dit, tous les modèles individuels ont une structure identique avec des équations et des variables semblables mais des paramètres propres à chaque ménage. Chaque modèle individuel est résolu séparément, excepté dans la phase d'estimation des paramètres de la fonction comportementale ou dans le cas où il pourrait y exister un échange de facteurs de production entre les exploitations (ce qui n'est pas le cas dans cette étude). Dans ces deux cas, toutes les exploitations individuelles de chaque région sont optimisées simultanément.

Dans le cas du Sénégal, le module de consommation a été, cependant, désactivé en raison du manque des données sur la consommation de référence des ménages sondés, réduisant ainsi FSSIM-Dev à un modèle d'offre agricole. Outre l'utilisation de la méthode bayésienne HPD, comme indiqué ci-dessus, le modèle a été calibré en utilisant des informations préalables sur les élasticités de l'offre (Louhichi et al. 2018). Les paramètres de la fonction comportementale ont été estimés de sorte que le modèle reproduise exactement l'assolement observé et génère des élasticités de l'offre similaires à celles observées dans la littérature. Le calibrage est effectué d'une manière non-myope, c'est-à-dire en prenant en compte les effets du changement du profit marginal des activités non seulement sur l'assolement mais aussi sur les valeurs marginales des contraintes ressources (Heckelei 2002, Mérel and Bucaram 2010). Les paramètres de la fonction comportementale sont estimés uniquement pour les activités observées dans chaque ménage agricole, ce qui signifie que le problème d'auto-sélection n'est pas explicitement traité dans cette estimation. Pour résoudre ce problème, nous avons adopté l'hypothèse suivante : la marge brute des activités non-observées (alternatives) est égale à la marge brute moyenne du groupe, le terme quadratique de la fonction comportementale est égal au terme moyen du groupe et le terme linéaire de la fonction comportementale est dérivé de la différence entre la marge brute et les valeurs marginales des contraintes ressources. Autrement dit, l'adoption de nouvelles activités (non-observées) par une exploitation reste possible en utilisant les paramètres des activités observées dans d'autres exploitations du même type.

# 4 Application du modèle FSSIM-Dev au Sénégal

## 4.1 Données utilisées

L´application du modèle FSSIM-Dev pour le cas du Sénégal est basée sur les données issues de la seconde Enquête de Suivi de la Pauvreté au Sénégal 2 (ESPS2) réalisée en 2011 (une première enquête avait déjà été réalisée en 2005) couvrant l'ensemble du pays. Cette enquête fait appel à une méthode d'échantillonnage par grappes à deux degrés avec stratification au premier degré. Les unités statistiques, ou unités primaire sont les districts de recensement. 337 districts de recensement ont été ainsi tirés sur un nombre total de 9780 dans tout le pays (ces districts correspondent à un ou plusieurs villages). Au second degré, dans chacun des districts tirés, 18 ménages ont été interrogés. Ce sont au total 5953 ménages représentatifs de l'ensemble du pays qui ont été enquêtés.

L'enquête comporte 2 modules :

i) un premier module portant sur les membres du ménage et renseignant les caractéristiques individuelles, le niveau d'éducation, la santé, l'emploi, et les migrations au sein du ménage.

ii) un second module portant sur le ménage et renseignant les caractéristiques du logement, les accès aux services sociaux, les sources de revenu, et enfin les activités agricoles détaillées par culture (mil, sorgho, riz, mais, riz, arachide…).

Les informations utilisées dans cette étude portent sur :

— Les surfaces emblavées et les niveaux de production
— Les quantités d'intrants utilisés (semences, engrais, produit phytosanitaires, matériel agricole)
— L'utilisation de la main d'œuvre familiale et salariée
— Les prix des cultures et des intrants
— Les amortissements du matériel agricole

Avant d'intégrer ces variables dans le modèle, plusieurs étapes ont été nécessaires: construction de certaines variables clés du modèle à partir des variables du questionnaire, mise en forme adéquate des données et enfin nettoyage des données par la méthode de Tuckey basée sur l'intervalle interquartile ou par windsoring. Les valeurs manquantes étaient remplacées par la médiane.

Après sélection des seuls ménages agricoles et une fois le nettoyage des données effectué, ce sont 2 278 ménages agricoles qui sont contenus dans la base de données finale sur lesquels sont réalisés les simulations.

## 4.2 Statistiques descriptives

Comme mentionné plus haut, l'activité agricole au Sénégal se déroule principalement durant la saison des pluies, aussi appelée hivernage, qui s'étend du mois de juin au mois de septembre/octobre. Néanmoins, certaines cultures sont pratiquées durant le reste de l'année, lors de la période sèche, à condition de disposer d'infrastructures d'irrigation. Les cultures pratiquées à ce moment de l'année sont celles à forte valeur-ajoutées, principalement les cultures maraîchères. Dans notre base de données, les cultures pratiquées durant la période sèche représentent moins de 3% des observations et moins de 1.5% des surfaces totales emblavées. De plus, il s'agit seulement des cultures d'aubergines, oignon et tomates. Plutôt que de distinguer ces deux périodes dans la présentation des statistiques descriptives et des résultats, elles seront fusionnées afin de faciliter la lecture des résultats et des commentaires associés. Notons néanmoins que dans la modélisation, ces deux périodes ont été bien distinguées. Cela permet de considérer la contrainte de terre et de travail puisque certaines productions ne sont produites que pendant l'hivernage alors que d'autres seulement durant la saison sèche.

Le Tableau 3 présente les caractéristiques principales des variables d'intérêt aux niveaux régional et national. D'après ce tableau, on constate que la taille moyenne des exploitations agricoles est assez homogène entre les régions sénégalaises (autour de 5 ha), à l'exception de la capitale (Dakar) et de la zone de la vallée du fleuve Sénégal (St Louis et Matam) où elle est beaucoup plus petite.

Par ailleurs, les statistiques sur les assolements observés retranscrivent bien les caractéristiques principales des systèmes de production agricoles sénégalais déjà connus. On retrouve par exemple la culture de riz pratiquée dans les régions de St-Louis, de la Casamance et dans une moindre mesure de Matam. Pour les autres régions, on retrouve des assolements qui comprennent pour près de la moitié des surfaces une céréale sèche (principalement mil et sorgho, mais aussi maïs) éventuellement complémenté par des tubercules (manioc et patate douce) comme dans le cas de Thiès et pour l'autre moitié des cultures de rente (principalement arachide et niébé mais aussi dans une moindre mesure coton, bissap, noix de cajou et enfin des produits maraichers). Enfin, notons que les statistiques reproduisent aussi l'importance des productions horticoles dans les régions de St Louis et de Thiès. Cette dernière région contenant la région des Niayes connue pour approvisionner l'agglomération dakaroise en cultures maraîchères.

Le tableau 3 présente aussi les taux d'utilisation de l'engrais, les quantités appliquées par ha et enfin les taux de bénéficiaires du programme de subvention des intrants (engrais). Si en moyenne 38.3% des ménages sénégalais utilisent au moins 1 kg d'engrais, ce sont seulement 8.7 kg/ha d'engrais qui sont appliqués en moyenne, ce qui est assez faible mais conforme aux autres données existantes. Par exemple les données de la banque mondiale indiquent pour l'année 2011 une moyenne de 8.2 kg d'engrais par ha en moyenne. Il existe néanmoins de fortes disparités entre les régions. La région dakaroise et de St Louis sont celles qui ont le plus recours aux engrais avec respectivement 62% et 75% des ménages qui utilisent de l'engrais et en moyenne 31.4 kg/ha et 34.1 kg/ha d'engrais appliqués. Dans les régions du bassin arachidier ce sont autour de 30% des ménages qui ont recours aux engrais, avec en moyenne 9 kg/ha utilisés. En Casamance, ce sont 43% des ménages qui ont recours à l'engrais ce qui est assez élevé, mais la quantité moyenne utilisée est très faible, d'environ 6 kg/ha. Enfin, dans les régions de l'Est du pays plus reculé (Louga, Matam, Tambacounda, Kédougou) les quantités moyennes d'engrais utilisées à l'hectare sont extrêmement réduites (autour de 3 kg/ha), avec des taux d'utilisation et des taux de bénéficiaires eux-aussi inferieures au aux autres régions du pays. De manière globale, on note des taux d'utilisation relativement corrects associés à des quantités à l'hectare très faibles qui s'explique par le programme de subvention. En effet, hormis pour Dakar et St Louis, les taux de bénéficiaire du programme de subvention des intrants parmi les utilisateurs d'engrais sont proches de 90% (légèrement plus faibles pour la Casamance et Tambacounda/Kédougou). Sans ce programme, il est fort probable que de nombreux agriculteurs n'utiliseraient pas d'engrais.

## 4.3   Typologie des pratiques culturales

Afin de prendre en compte l'hétérogénéité des pratiques cultures et ainsi distinguer différents niveaux d'intensification dans le modèle, une typologie des pratiques culturales a été élaborée pour les cultures principales en utilisant la méthode des classifications ascendantes hiérarchiques (CAH). Les CAH permettent de regrouper les observations de l'échantillon deux à deux jusqu'à n'obtenir qu'un seul groupe. Les variables retenues pour ces classifications sont les dépenses par ha pour les postes budgétaires suivants :

— Semence

— Engrais

— Produit phytosanitaire

— Matériel agricole (dépenses d'entretien et/ou location)

— Embauche de main d'œuvre extérieure

Pour chacune des cultures principales, deux types de pratiques ont pu être distinguées : extensive (ext) et semi-intensive (semi-int). Les principales caractéristiques de ces pratiques sont exposées dans le Tableau 4. Notons que les cultures principales représentent 84% des observations et surtout 88% des surfaces totales cultivées.

Si on compare le nombre d'observations classées dans la catégorie ext avec ceux classés dans la catégorie semi-int, ce sont en moyenne autour de 20% des observations qui appartiennent à la seconde catégorie (semi-int). Ce taux varie entre 13% pour le maïs et 23% pour l'arachide. On constate que, comme attendu, la pratique semi-int aboutit à des dépenses en intrants significativement plus élevées que celles des pratiques ext, variant entre +400% pour l'arachide et +800% pour le riz (les écarts sont très élevés car les dépenses dans le cas de la technique extensive sont très faibles). Ces hausses des dépenses en intrants sont accompagnées par des hausses de rendements compris entre +35% (arachide) et +110% (riz) et par un impact positif sur les marges brutes puisqu'elles sont toutes supérieures pour la techniques semi-intensive. Les augmentations des marges brutes vont de +7% pour l'arachide et le mil à +95% pour le riz. Il semble ainsi économiquement justifié d'intensifier les cultures, du moins celles pour lesquelles les analyses ont été réalisées.

Parmi les cultures qui dégagent la plus forte marge brute on trouve le riz et le manioc, suivi respectivement par le maïs, l'arachide et le sorgho (en semi-intensif, le sorgho dégage une marge brute plus grande que l'arachide). Le mil ferme la marche.

## 4.4 Typologie des exploitations agricoles

Pour tenir compte de la forte hétérogénéité des exploitations agricoles au Sénégal mais aussi pour agréger et présenter aisément les résultats du modèle, nous avons décidé de classer les exploitations en groupes homogènes selon deux critères :

— La taille économique

— La spécialisation

La taille économique d'une exploitation est mesurée par sa valeur économique totale qui correspond à la somme des valeurs économiques de toutes les productions végétales de l'exploitation (production totale d'un produit multiplié par son prix). Une fois l'ensemble des tailles économiques mesurées, les exploitations ont été réparties en trois groupes selon leurs valeurs obtenues : petites, moyennes, et grandes exploitations (Tableau 5).

En se basant sur ce critère, on constate, d'une part, une dominance des petites exploitations dans toutes les régions et d'autre part, une forte concentration des grandes exploitations dans la région dakaroise, de Kaffrine et de St Louis. Ce sont aussi la région de Dakar et de St Louis qui abritent une grande proportion d'exploitation de petite taille économique (ce sont en fait les exploitations moyennes qui sont peu représentées dans ces deux régions), tout comme les régions de Matam et de Casamance.

La spécialisation consiste à classer les exploitations selon leur production principale (par exemple « céréalières »). La catégorisation des exploitations selon leur spécialisation a été réalisée comme suivant. Tout d'abord, nous avons identifié les 4 principaux types de culture à savoir :

— Les céréales (mil, sorgho, riz...)

— Les racines et tubercules (manioc, patate douce...)

— Les légumineuses (arachide, niébé...)

— Les cultures de rente : cultures maraichères (tomate, aubergine, pastèque...) et autres (coton, anacardier, ...)

Ensuite pour chaque exploitation, nous avons calculé la part de chaque type de culture dans la valeur économique totale. Ainsi, une exploitation est classée comme spécialisée dans un (ou plusieurs) type(s) de culture si au moins 65% de sa valeur économique totale provenait de ce (ou ces) type(s) de culture. Dans le cas contraire, l'exploitation est classée comme mixte. Au total, ce sont 4 types d'exploitations qui ont été retenues (Tableau 6).

D´après ce tableau, on observe qu'un peu plus de la moitié des exploitations associent principalement une culture céréalière (mil, sorgho, maïs ou riz) avec une légumineuse (principalement l'arachide). Vient ensuite autour de 20% des exploitations qui cultivent principalement des cultures vivrières. Ce sont notamment des exploitations de petites tailles aussi bien physique (2.7 ha en moyenne) qu'économique (276.2 milliers de FCFA). En troisième position, on trouve les exploitations spécialisées dans les cultures de rente (14.7% de l'échantillon) avec une taille physique proche de la moyenne de l'échantillon (4.7 ha) et une taille économique plutôt importante. Enfin, 11.2% de l'échantillon est représentée par des exploitations qui ne sont pas spécialisées dans un type de culture en particulier. Elles ont donc une part à peu près égale entre culture vivrière et culture de rente. Elles ont une taille économique inferieure aux exploitations spécialisées dans les cultures de rente mais supérieure aux autres exploitations.

Le Tableau 7 indique que les exploitations spécialisées dans le vivrier sont basées notamment à St Louis, Matam, Casamance, Tambacounda et Koudougou. Celles associant une céréale à une légumineuse se retrouve principalement dans le bassin arachidier (Thiès, Diourbel, Fatick...). Celles spécialisées dans les cultures de rente sont plutôt à Dakar, St Louis, Louga et Thiès. Enfin, la Casamance regroupe presque le quart des exploitations dites mixtes.

**Tableau 3**: Caractéristiques principales de l'échantillon

|  | Sénégal | Dakar | Thiès | Diourbel | Fatick Kao | Kaffrine | St-Louis | Louga | Matam | Tamba Ked | Casamance |
|---|---|---|---|---|---|---|---|---|---|---|---|
| Nombre d´exploitations | 2278 | 8 | 118 | 136 | 392 | 240 | 141 | 65 | 163 | 437 | 578 |
| Surface moyenne d´exploitation (ha) | 4.7 | 3.5 | 5.1 | 5.6 | 4.9 | 7.6 | 3.5 | 5.2 | 2.5 | 4.1 | 4.6 |
| Nombre équivalent adulte | 5.1 | 6.9 | 6 | 5.5 | 5 | 5 | 5 | 5.1 | 4.5 | 5.3 | 5.2 |
| Assolement (% par région) | | | | | | | | | | | |
| Céréales (hors riz) | 43.9 | 31 | 30.8 | 45.3 | 52.2 | 50.3 | 15.8 | 40.6 | 68.2 | 59.3 | 38.2 |
| Riz | 7.1 | - | - | - | 0.7 | 0.1 | 31.7 | - | 4.7 | 1.9 | 22.9 |
| Tubercule | 2 | 10.9 | 8.9 | 1.1 | 0.2 | 0.4 | 6.4 | 0.3 | 1.7 | 0.2 | 1.1 |
| Cultures maraichères | 7.6 | 3.3 | 18.8 | 5.7 | 4.5 | 7 | 24.5 | 5.8 | 11.1 | 3.5 | 4 |
| Cultures de rente | 39.4 | 54.8 | 41.5 | 47.9 | 42.4 | 42.2 | 21.6 | 53.3 | 14.3 | 35.1 | 33.8 |
| Taux d´utilisation engrais (%) | 38.3 | 62.5 | 38.1 | 21.3 | 36.2 | 59.1 | 75.2 | 10.8 | 10.4 | 29.5 | 43.4 |
| Taux de bénéficiaire des subvention (%) | 31.3 | 50 | 34.7 | 19.9 | 31.3 | 52.1 | 46.8 | 9.2 | 7.3 | 23.3 | 35.9 |
| Taux de bénéficiaire parmi les utilisateurs (%) | 81.8 | 57.7 | 91.9 | 93.8 | 87.2 | 89.6 | 60.9 | 95.8 | 85.5 | 77.1 | 73.3 |
| Quantités d´engrais utilisées (kg/ha) | 8.7 | 31.4 | 9.1 | 6.9 | 7.6 | 11.2 | 34.1 | 0.7 | 2 | 3.5 | 6 |

*Source :* calculs des auteurs à partir des données ESPS2.

**Tableau 4**: Caractéristiques principales des pratiques culturales identifiées par Classification Ascendante Hiérarchique

| | Nombre d'observation | | Dép. semences[1] | | Dép. engrais | | Dép. phyto. produit | | Dép. matériel | | Dép. main d'œuvre | | Rendement (kg/ha) | | Prix de vente | Marge brute à l'hectare | |
|---|---|---|---|---|---|---|---|---|---|---|---|---|---|---|---|---|---|
| | Ext.[2] | Semi-int. | Ext. | Semi-int. | Ext. | Semi-int. | Ext. | Semi-int. | Ext. | Semi-int. | Ext. | Semi-int. | Ext. | Semi-int. | | Ext. | Semi-int. |
| Mil | 1 377 | 238 | 3 874 | 23 718 | 1 172 | 9471 | 369 | 1 317 | 973 | 2 794 | 720 | 6 476 | 604 | 854 | 173.8 | 97 850 | 104 580 |
| Sorgho | 250 | 68 | 1 829 | 14 138 | 659 | 6 888 | 333 | 1 000 | 1 148 | 2 182 | 141 | 7 343 | 666 | 1 005 | 191.4 | 123 267 | 160 883 |
| maïs | 801 | 115 | 2 528 | 20 428 | 1 354 | 9 938 | 450 | 2 296 | 759 | 1 859 | 800 | 7 899 | 752 | 1 110 | 194.6 | 140 526 | 173 566 |
| Riz | 418 | 91 | 3 301 | 20 509 | 1 603 | 20 799 | 535 | 4 974 | 569 | 4 279 | 1 045 | 13 417 | 1 095 | 2 372 | 224.8 | 239 103 | 469 248 |
| Manioc | 55 | 21 | 4 458 | 27 621 | 1 780 | 5 850 | 562 | 647 | 1 138 | 3 684 | 766 | 14 424 | 789 | 1 851 | 273.3 | 206 930 | 453 652 |
| Arachide | 1 156 | 347 | 4 443 | 27 769 | 1 668 | 7 881 | 555 | 1 213 | 1 105 | 3 312 | 1 232 | 5 462 | 721 | 975 | 185.4 | 124 578 | 135 053 |

[1] Dép = Depense (FCFA) ; [2] Ext = extensif et Semi-Int = semi-intensif Source : calculs des auteurs à partir des données ESPS2.

**Tableau 5**: Caractéristiques principales des exploitations selon leur taille économique et leurs répartitions régionales

| | Valeurs seuils des tailles économiques (en FCFA) | Distribution des classes selon les régions (en %) | | | | | | | | | | |
|---|---|---|---|---|---|---|---|---|---|---|---|---|
| | | Sénégal | Dakar | Thiès | Diourbel | Fatick_Kao | Kaffrine | St-Louis | Louga | Matam | Tamba_Ked | Casamance |
| Petite exploitation | Moins de 400 000 | 47.45 | 62.5 | 49.15 | 45.59 | 41.84 | 20.42 | 51.77 | 46.15 | 84.05 | 47.14 | 51.38 |
| Moyenne exploitation | Entre 400 et 850 000 | 30.43 | 0 | 27.12 | 31.62 | 39.03 | 34.58 | 15.6 | 35.38 | 12.27 | 39.13 | 25.26 |
| Grande exploitation | Plus de 850 000 | 22.12 | 37.5 | 23.73 | 22.79 | 19.13 | 45 | 32.62 | 18.46 | 3.68 | 13.73 | 23.36 |

Source: calculs des auteurs à partir des données ESPS2

**Tableau 6**: Caractéristiques principales des exploitations selon leur spécialisation

| Exploitations spécialisées en : | Description | Part dans l'échantillon | Surface moyenne (ha) | Taille économique moyenne (FCFA) |
|---|---|---|---|---|
| Culture vivrière | Spécialisation dans la production de céréales et/ou racines et/ou tubercules | 19.9% | 2.7 | 276 259 |
| Culture de rente | Spécialisation dans la production de légumineuses et/ou de cultures de rente | 14.7% | 4.5 | 1 025 415 |
| Céréales/légumineuses | Spécialisation dans la production de céréales ou de légumineuses | 54.2% | 5.8 | 666 161 |
| Mixte | Non spécialisée | 11.2% | 6.5 | 907 493 |

Source: calculs des auteurs à partir des données ESPS2

**Tableau 7**: Distribution régionales des exploitations selon leur spécialisation (en pourcentage)

|  | Cultures vivrières | Cultures de rente | Céréales/légumineuses | Mixte |
|---|---|---|---|---|
| Sénégal | 19.9 | 14.7 | 54.2 | 11.2 |
| Dakar | 20 | 66.67 | 13.33 | 0 |
| Thiès | 5.49 | 22.54 | 61.56 | 10.4 |
| Diourbel | 8.33 | 10.61 | 81.06 | 0 |
| Fatick_Kao | 12.35 | 7.37 | 77.6 | 2.67 |
| Kaffrine | 11.92 | 10.54 | 74.03 | 3.51 |
| St-Louis | 30.09 | 56.43 | 10.03 | 3.45 |
| Louga | 8.59 | 36.87 | 54.55 | 0 |
| Matam | 59.66 | 11.08 | 17.05 | 12.22 |
| Tamba_Ked | 20.23 | 10.87 | 55.08 | 13.82 |
| Casamance | 25.03 | 13.49 | 38.29 | 23.19 |

Source: calculs des auteurs à partir des données ESPS2

## 4.5 Scenarios simulés

Le calibrage du modèle a été réalisé pour l'année 2011 en utilisant les données collectées auprès de 2 278 ménages agricoles sénégalais. Comme expliqué plus haut, le calibrage correspond à l'estimation des paramètres de la fonction de comportement de chaque exploitation de manière à ce que les assolements observés cette année-là soient parfaitement simulés par le modèle. Les résultats du modèle une fois calibré correspondent à l'année de base (*baseyear*) qui est donc ici 2011. Toutefois, depuis 2011, le contexte économique a sensiblement évolué au Sénégal. Aussi, plutôt que de comparer les résultats des différents scenarios avec cette année de base, il est préférable de les comparer avec une situation de référence plus récente (l'année 2017) et proche de celle à laquelle sont confrontés les ménages agricoles actuellement. Cette situation de référence, appelée aussi *baseline*, servira de base de comparaison à l'évaluation de l'impact des mesures simulées.

En comparaison à l'année de base, il a été assumé dans la *baseline* que (i) les coûts de production ont progressé au même rythme que l'inflation (un taux d'inflation de 2.7% par an est ainsi retenu (Banque Mondiale, 2020), (ii) les rendements et les prix des cultures ont subi une variation similaire que celle enregistrée au niveau national entre 2011 et 2017 et reportée dans les statistiques officielles (Tableau 8) et (iii) la politique de subventions aux intrants reste inchangée entre 2011 et 2017.

**Tableau 8** : Variation des rendements et des prix des principales cultures entre la baseyear et la baseline (en pourcentage)

|  | Variations des rendements (%) | Variations des prix (%) |
|---|---|---|
| Arachide | 34.1 | 13.5 |
| Mais | 33.2 | 9.8 |
| Mil | 13.5 | 24.0 |
| Oignon | 4.7 | 30.4 |
| Riz paddy | 13.3 | 29.4 |

Source: prix: Direction des Prévisions et des Statistiques du Ministère de l'Economie, du Plan et de la Coopération; rendements: FAOSTAT.

Nous allons maintenant décrire les trois scénarios qui seront implémentés et comparés à la situation de référence (*baseline*).

### 4.5.1 Scenario 1 : abolition du programme de subvention aux engrais actuel

Ce premier scenario a pour but d'estimer les effets du programme de subvention actuellement en place au Sénégal. Pour ce faire, on va supposer que ce programme est totalement suspendu et par conséquent les bénéficiaires du programme doivent maintenant acheter leur engrais au prix du marché. Il s'agit donc d'un scénario "ex-post" où l'impact du programme correspondra à l'inverse de la différence entre le scénario et la *baseline*. Dans la suite du texte, ce scénario est appelé *Abol*, par référence à l'abolition des subventions.

### 4.5.2 Scénario 2 : mise en place d'un programme de subvention aux engrais universel plafonné

Le programme de subvention des engrais actuellement en place se veut accessible à tous, c'est-à-dire *universel*. Néanmoins, comme on l'a vu précédemment, pour des raisons budgétaires, les engrais subventionnés ne sont pas disponibles pour l'ensemble des ménages agricoles qui le désirent. Une sélection des exploitations de facto se fait de sorte qu'il y a des bénéficiaires et des non-bénéficiaires du programme. Dans ce scénario, on suppose qu'il n'y a pas de contrainte budgétaire et que le gouvernement est dans la mesure d'appliquer un programme réellement universel. L'ensemble des ménages agricoles peuvent donc bénéficier des engrais à un prix subventionné s'il le souhaite. Un plafond de 75 kg d'engrais par ménage est, néanmoins, appliqué, ce qui correspond à la moitié du quota existant dans le programme en cours. L'idée donc d'étendre le programme à l'ensemble des ménages tout en restant raisonnable en terme de budget et de disponibilité en engrais. Dans la suite du texte, ce scénario est appelé *Univ*, pour le caractère universel du programme qui est simulé.

### 4.5.3 Scenario 3 : mise en place d'un programme de subvention aux engrais ciblé

Le troisième scenario correspond à un programme de subvention des intrants *ciblé*, ou seulement les exploitations de 5 ha ou moins peuvent bénéficier de la subvention. L'objectif étant de faciliter l'accès aux engrais uniquement pour les exploitations de tailles petites et moyennes. Avec une taille moyenne des exploitations dans notre échantillon de 4.7 ha et une médiane qui se situe à 3.8 ha, se sont plus de la moitié qui pourraient potentiellement bénéficier de la mesure. De plus, Un plafond de 75 kg d'engrais par ménage est aussi appliqué dans ce scénario. Puisqu'une partie seulement des ménages est ciblée par le programme, ce scénario est appelé *Cibl*.

Le Tableau 9 présente le pourcentage de bénéficiaire potentiel du programme, sous chaque scénario, pour l'ensemble du pays et ses régions. Si les pourcentages sont évidents pour les deux premiers scenarios, le tableau permet de vérifier que le troisième scenario aboutit à des taux de bénéficiaires supérieurs à ceux du scenario de référence (*baseline*).

**Tableau 9**: Taux de bénéficiaire des subventions aux engrais sous chaque scenario (en pourcentage)

|  | Baseline | Abol (scénario 1) | Univ (scénario 2) | Cibl (scénario 3) |
|---|---|---|---|---|
| Sénégal | 31.3 | 0 | 100 | 62 |
| Dakar | 50 | 0 | 100 | 87.5 |
| Thiès | 34.7 | 0 | 100 | 61.8 |
| Diourbel | 19.9 | 0 | 100 | 52.9 |
| Fatick_Kao | 31.3 | 0 | 100 | 57.4 |
| Kaffrine | 52.1 | 0 | 100 | 32.5 |
| St-Louis | 46.8 | 0 | 100 | 80.8 |
| Louga | 9.2 | 0 | 100 | 53.8 |
| Matam | 7.3 | 0 | 100 | 88.9 |
| Tamba_Ked | 23.3 | 0 | 100 | 66.8 |
| Casamance | 35.9 | 0 | 100 | 62.9 |

Source : calculs des auteurs à partir des données ESPS2

Les effets de ces trois scénarios sont maintenant présentés et analysés en les comparant à la baseline. L'analyse des résultats portera essentiellement sur l'utilisation des terres, la production, les revenus agricoles et enfin le coût budgétaire de ces scenarios.

# 5 Résultats et discussions

Afin de faciliter la présentation et la comparaison des résultats de différents scénarios, nous présentons, en premier lieu, les résultats, en valeurs absolues, du scenario de référence (baseline) et, ensuite, nous exposons les effets, en termes relatifs, des trois scénarios simulés par rapport à cette baseline. Il faut noter que les résultats de la baseline sont différents de ceux de l'année de base présentés dans le tableau 4 du fait des projections des prix et des rendements introduites entre l'année de base et la baseline mais aussi et surtout parce que les résultats de la baseline sont pondérés par le poids de chaque exploitation pour agréger les moyennes aux niveaux régionaux/national, ce qui n'est pas le cas pour les statistiques descriptives présentées dans le tableau 4 qui sont des moyennes de l'échantillon sans pondération. Les poids statistiques utilisés proviennent directement de l'enquête ESPS-2.

## 5.1 Résultats du scenario de référence (baseline)

Le Tableau 10 donne les surfaces totales cultivées, les volumes produits, les rendements ainsi que les quantités moyennes d'engrais utilisés à l'hectare pour les principales cultures obtenus dans le scenario de référence. On remarquera que, comme prévu, les engrais sont utilisés en premier lieu pour les céréales à savoir le maïs et le riz, puis pour le coton et les productions horticoles (pastèque, oignon). Les rendements du mil, sorgho, maïs et riz sont en lignes avec ceux que l'on retrouve dans la littérature et en particulier dans la base de données FAOSTAT pour l'année 2017. Dans le cas du riz, la région de St Louis obtient des rendements bien plus élevés que dans le reste du pays (notamment la Casamance) car c'est la seule région du pays ou les systèmes de production rizicole irrigués sont développés à grandes échelles. On note aussi une forte disparité régionale au niveau de l'application d'intrants. Par exemple, si on prend le cas du mil, 9 kg/ha d'engrais sont utilisés en moyenne dans les régions de Kaffrine et St Louis, alors que dans les régions de Louga et Tambakounda/Kedougou, seulement 1 kg/ha d'engrais en moyenne sont utilisés. Aussi, on observe que certaines productions sont cultivées dans quelques régions exclusivement. C'est le cas du riz qui nécessite de conditions de production bien particulières mais aussi de la noix de cajou, du coton, et de l'oignon. De même, la pastèque, est principalement cultivée dans le bassin arachidier.

Le Source: résultats du modèle

**Tableau 11** présente les mêmes résultats, mais en fonction de la taille économique des exploitations. On peut s'apercevoir à partir des surfaces que certaines cultures sont surtout l'apanage des grands producteurs. Il s'agit du coton et des cultures horticoles (pastèque, oignon et aubergine). Ce sont des cultures de rente qui nécessite des intrants et des débouchés auxquelles n'ont pas toujours accès les petits producteurs. Par ailleurs, on s'aperçoit que généralement, les rendements augmentent avec la taille économique des producteurs, tout comme les quantités d'engrais utilisées. Par exemple pour le coton, les petits producteurs utilisent en moyenne 6.6 kg/ha d'engrais pour un rendement moyen de 658 kg/ha alors que les grandes exploitations utilisent en moyenne 16 kg/ha pour un rendement moyen de 1291 kg/ha.

Le Tableau 12 indique les même résultats selon la spécialisation des exploitations agricoles. Il montre que les rendements des cultures céréalières sont légèrement supérieurs pour les exploitations spécialisées en culture de rente et les exploitations mixtes par rapport aux exploitations vivrières ou spécialisées en céréales-légumineuses. Dans le cas du mil, les rendements pour les premières exploitations sont de 1031 kg/ha et 1186 kg/ha alors que pour les dernières exploitations, les rendements sont de 905 kg/ha et 987 kg/ha. Pourtant, en termes d'engrais utilisés, aucune tendance ne se dégage véritablement selon la spécialisation des exploitations, hormis que les exploitations spécialisées en cultures vivrières utilisent moins d'intrants que les autres exploitations (les fortes valeurs pour les cultures d'oignon et aubergine et liés au très faible nombre d'observations d'exploitations spécialisés en cultures vivrières qui cultivent ces deux cultures, ce qui rend les valeurs moyennes très sensibles à des valeurs extrêmes sans qu'elles ne soient des outliers). En volume, les céréales sont avant tout produites par les exploitations spécialisées en céréales-légumineuses à l'exception du riz dont la principale source de production provient des exploitations vivrières. Logiquement, ce sont les exploitations spécialisées dans les cultures de rente qui produisent l'essentiel des cultures horticoles.

Enfin, les Tableau 13 et Tableau 14 présentent les revenus annuels moyens des exploitations selon leur localisation, leur taille économique et leur spécialisation. Le revenu annuel moyen des exploitations à l'échelle du pays est de 1 386 438 FCFA (2 113 €), correspondant à un revenu par jour de 3 798 FCFA, soit 5.7 €[1].

---
[1] 1 euro = 655 FCFA

Cependant cette moyenne masque des disparités importantes entre les régions ayant un fort potentiel agricole comme les régions de St Louis ou de Kaffrine (revenus annuels moyens supérieurs à 2 000 000 FCFA, soit 3 048 €) et des régions beaucoup plus pauvres telles que les régions de Matam avec un revenu annuel moyen des exploitations de 404 379 FCFA (616 €) ou les régions de Fatick et Kaolack avec un revenu moyen de 970 203 FCFA (1 478 €). Par ailleurs, comme attendu, le revenu annuel moyen des exploitations augment avec leur taille économique. Les petites exploitations ont un revenu annuel moyen de 411 893 FCFA (627 €) alors que les grandes exploitations ont un revenu moyen de 3 316 865 FCFA (5 050 €). Enfin, le revenu des exploitations spécialisées en cultures de rente (2 325 430 FCFA, soit 3540 €) et les exploitations dites mixtes (2 135 574 FCFA, soit 3 255 €) ont de revenus significativement supérieurs aux exploitations vivrières (757 875 FCFA, soit 1 155 €) et aux exploitations spécialisées dans les cultures céréalières et légumineuses (1 287 355 FCFA, soit 1 962 €).

**Tableau 10**: Variables de production (surface, production, rendement, application d'engrais) dans la baseline pour les principales cultures et selon les régions

| | | Sénégal | Dakar | Thiès | Diourbel | Fatick_Kao | Kaffrine | St-Louis | Louga | Matam | Tamba_Ked | Casamance |
|---|---|---|---|---|---|---|---|---|---|---|---|---|
| Mil | Surface (ha) | 658 840 | 3 399 | 60 150 | 108 546 | 194 078 | 77 197 | 4 459 | 34 411 | 29 140 | 35 317 | 112 143 |
| | Production (t) | 644 109 | 4 587 | 45 598 | 95 178 | 180 166 | 90 646 | 6 152 | 30 086 | 22 712 | 34 876 | 134 108 |
| | Rendement (kg/ha) | 978 | 1 350 | 758 | 877 | 928 | 1 174 | 1 380 | 874 | 779 | 988 | 1 196 |
| | Qté engrais (kg/ha) | 6.4 | 75.6 | 2.9 | 5.6 | 7.9 | 9.4 | 9.1 | 0.5 | 0.3 | 1.5 | 7.2 |
| Sorgho | Surface | 79 642 | 0 | 7 864 | 2 038 | 10 701 | 11 770 | 9 650 | 1 636 | 2 995 | 20 166 | 12 822 |
| | Production | 91 527 | | 7 543 | 1 104 | 10 930 | 13 324 | 11 651 | 1 265 | 2 923 | 25 860 | 16 927 |
| | Rendement | 1 149 | | 959 | 542 | 1 021 | 1 132 | 1 207 | 773 | 976 | 1 282 | 1 320 |
| | Qté engrais | 4.5 | | 1.3 | 3.5 | 3.3 | 8.8 | 4.1 | 0 | 0.2 | 2.5 | 8.7 |
| Maïs | Surface | 193 984 | 0 | 4 878 | 2 887 | 26 350 | 20 475 | 5 128 | 6 850 | 7 145 | 50 534 | 69 737 |
| | Production | 232 822 | | 9 026 | 3 157 | 29 537 | 23 878 | 8 306 | 6 080 | 5 012 | 60 327 | 87 497 |
| | Rendement | 1 200 | | 1 850 | 1 094 | 1 121 | 1 166 | 1 620 | 888 | 702 | 1 194 | 1 255 |
| | Qté engrais | 7.7 | | 10 | 0 | 15.2 | 14.8 | 13.6 | 0.4 | 0.5 | 2.9 | 7.6 |
| Riz | Surface | 173 525 | 0 | 0 | 0 | 4 123 | 51 | 33 712 | 0 | 2 845 | 3 629 | 129 166 |
| | Production | 238 916 | | | | 2 931 | 41 | 80 234 | | 2 451 | 3 906 | 149 351 |
| | Rendement | 1 377 | | | | 711 | 811 | 2 380 | | 862 | 1 076 | 1 156 |
| | Qté engrais | 10.3 | | | | 2.4 | 2.1 | 38.5 | | 18.5 | 7.7 | 3.1 |

Source: résultats du modèle

Tableau 10 (suite)

| | | Sénégal | Dakar | Thiès | Diourbel | Fatick_Kao | Kaffrine | St-Louis | Louga | Matam | Tamba_Ked | Casamance |
|---|---|---|---|---|---|---|---|---|---|---|---|---|
| Manioc | Surface (ha) | 30 390 | 1 772 | 21 095 | 2 869 | 451 | 834 | 211 | 306 | 0 | 79 | 2 774 |
| | Production (t) | 31 899 | 3 998 | 22 049 | 1 583 | 427 | 870 | 388 | 144 | | 88 | 2 352 |
| | Rendement (kg/ha) | 1 050 | 2 256 | 1 045 | 552 | 947 | 1 043 | 1 839 | 470 | | 1 114 | 848 |
| | Qté engrais (kg/ha) | 9.5 | 9.8 | 11.3 | 0 | 4.7 | 17.5 | 12.2 | 0 | | 35.5 | 3.5 |
| Arachide | Surface | 911 081 | 6 746 | 109 896 | 151 819 | 238 787 | 113 896 | 7 249 | 44 812 | 4 617 | 83 787 | 149 471 |
| | Production | 935 130 | 9 104 | 86 135 | 138 851 | 227 201 | 135 613 | 11 702 | 42 843 | 3 007 | 93 628 | 187 046 |
| | Rendement | 1 026 | 1 350 | 784 | 915 | 951 | 1 191 | 1 614 | 956 | 651 | 1 117 | 1 251 |
| | Qté engrais | 9 | 40.6 | 4.2 | 9 | 11.5 | 13.6 | 13.5 | 2.2 | 0.1 | 4.2 | 8.2 |
| Niébé | Surface | 155 198 | 1 346 | 36 283 | 27 847 | 22 338 | 9 054 | 16 479 | 21 402 | 4 557 | 2 605 | 13 287 |
| | Production | 142 361 | 2 892 | 30 663 | 21 303 | 16 862 | 10 734 | 20 006 | 18 913 | 3 602 | 1 893 | 15 492 |
| | Rendement | 917 | 2 149 | 845 | 765 | 755 | 1 186 | 1 214 | 884 | 790 | 727 | 1 166 |
| | Qté engrais | 2.4 | 39.6 | 1.8 | 1.5 | 1.8 | 4.2 | 4.9 | 0.4 | 0.1 | 6.6 | 2.2 |
| Noix de cajou | Surface | 20 099 | 0 | 0 | 0 | 0 | 0 | 0 | 0 | 0 | 0 | 20 099 |
| | Production | 23 045 | | | | | | | | | | 23 045 |
| | Rendement | 1 147 | | | | | | | | | | 1 147 |
| | Qté engrais | 3 | | | | | | | | | | 3 |

Source: résultats du modèle

Tableau 10 (suite)

|  |  | Sénégal | Dakar | Thiès | Diourbel | Fatick_Kao | Kaffrine | St-Louis | Louga | Matam | Tamba_Ked | Casamance |
|---|---|---|---|---|---|---|---|---|---|---|---|---|
| Coton | Surface (ha) | 18 662 | 0 | 0 | 0 | 0 | 605 | 0 | 0 | 0 | 4 159 | 13 898 |
|  | Production (t) | 22 011 |  |  |  |  | 555 |  |  |  | 4 768 | 16 689 |
|  | Rendement (kg/ha) | 1 179 |  |  |  |  | 917 |  |  |  | 1 146 | 1 201 |
|  | Qté engrais (kg/ha) | 16.7 |  |  |  |  | 16.3 |  |  |  | 6 | 19.9 |
| Pastèque | Surface | 16 010 | 0 | 458 | 2 035 | 2 487 | 5 256 | 3 164 | 874 | 91 | 257 | 1 389 |
|  | Production | 19 599 |  | 328 | 1 974 | 3 189 | 6 758 | 4 635 | 955 | 40 | 194 | 1 526 |
|  | Rendement | 1 224 |  | 716 | 970 | 1 282 | 1 286 | 1 465 | 1 093 | 443 | 755 | 1 099 |
|  | Qté engrais | 14 |  | 20 | 16 | 14 | 16 | 11 | 0 | 0 | 5 | 23 |
| Oignon | Surface | 15 814 | 0 | 3 858 | 0 | 827 | 33 | 9 404 | 0 | 461 | 43 | 1 188 |
|  | Production | 68 861 |  | 16 543 |  | 1 804 | 80 | 46 089 |  | 1 211 | 95 | 3 039 |
|  | Rendement | 4 354 |  | 4 288 |  | 2 181 | 2 428 | 4 901 |  | 2 626 | 2 213 | 2 558 |
|  | Qté engrais | 40.8 |  | 24 |  | 0 | 28.6 | 54.8 |  | 21.8 | 42.2 | 20.3 |
| Aubergine | Surface | 15 293 | 148 | 7 315 | 0 | 3 948 | 421 | 415 | 840 | 0 | 395 | 1 810 |
|  | Production | 19 259 | 50 | 16 195 |  | 1 283 | 150 | 377 | 283 |  | 128 | 793 |
|  | Rendement | 1 259 | 337 | 2 214 |  | 325 | 357 | 908 | 337 |  | 323 | 438 |
|  | Qté engrais | 11.9 | 0 | 12.6 |  | 11 | 4.7 | 71.2 | 0 |  | 5.3 | 6.7 |

Source: résultats du modèle

**Tableau 11**: Variables de production (surface, production, rendement, application d'engrais) pour la baseline selon la taille économique des exploitations

| | Surface (ha) | | | Production (t) | | | Rendement (kg/ha) | | | Quantité engrais (kg/ha) | | |
|---|---|---|---|---|---|---|---|---|---|---|---|---|
| | Petite | Moyenne | Grande | Petite | Moyenne | Grande | Petite | Moyenne | Grande | Petite | Moyenne | Grande |
| Mil | 152 450 | 229 285 | 277 106 | 122 587 | 204 584 | 314 534 | 804 | 892 | 1 135 | 6.3 | 5 | 8.8 |
| Sorgho | 14 830 | 29 795 | 35 018 | 11 883 | 31 174 | 45 922 | 801 | 1 046 | 1 311 | 3.2 | 3.5 | 5.9 |
| Maïs | 37 263 | 66 463 | 90 257 | 35 729 | 77 284 | 115 297 | 959 | 1 163 | 1 277 | 10.4 | 4.9 | 8.8 |
| Riz | 58 435 | 54 843 | 60 247 | 69 566 | 62 270 | 85 611 | 1 190 | 1 135 | 1 421 | 15.8 | 4.9 | 9.9 |
| Manioc | 5 287 | 11 390 | 13 713 | 6 445 | 10 292 | 15 419 | 1 219 | 904 | 1 124 | 15.9 | 6.1 | 9.1 |
| Arachide | 107 357 | 296 645 | 507 080 | 96 025 | 276 415 | 586 980 | 894 | 932 | 1 158 | 6.8 | 5.6 | 11.4 |
| Niébé | 42 086 | 48 197 | 64 916 | 29 805 | 42 548 | 65 533 | 708 | 883 | 1 010 | 1.8 | 2.6 | 2.8 |
| Noix de cajou | 1 415 | 3 524 | 15 160 | 1 122 | 3 493 | 18 585 | 793 | 991 | 1 226 | 2.3 | 2.5 | 4.4 |
| Coton | 44 | 4 938 | 13 680 | 29 | 4 396 | 17 664 | 658 | 890 | 1 291 | 6.6 | 11 | 16 |
| Pastèque | 115 | 133 | 15 762 | 132 | 92 | 19 144 | 1 151 | 694 | 1 215 | 13.6 | 34.3 | 13.9 |
| Oignon | 198 | 1 199 | 14 417 | 547 | 3 754 | 64 559 | 2 763 | 3 131 | 4 478 | 51.1 | 16.7 | 41.6 |
| Aubergine | 769 | 1 194 | 13 330 | 927 | 1 003 | 17 329 | 1 206 | 840 | 1 300 | 14.7 | 8.7 | 12.6 |

Source: résultats du modèle

**Tableau 12**: Variables de production (surface, production, rendement, application d'engrais) pour la baseline selon la spécialisation des exploitations

| | Surface (ha) | | | | Production (t) | | | | Rendement (kg/ha) | | | | Quantité engrais (kg/ha) | | | |
|---|---|---|---|---|---|---|---|---|---|---|---|---|---|---|---|---|
| | Vivrier | Rente | Ceré/leg | Mixte | Vivrier | Rente | Ceré/leg | Mixte | Vivrier | Rente | Ceré/leg | Mixte | Vivrier | Rente | Ceré/leg | Mixte |
| Mil | 136 450 | 41 849 | 438 483 | 42 058 | 123 541 | 43 139 | 432 885 | 49 878 | 905 | 1 031 | 987 | 1 186 | 3.4 | 10.3 | 6.1 | 7 |
| Sorgho | 28 501 | 2 480 | 42 335 | 6 326 | 32 418 | 2 866 | 48 140 | 7 862 | 1 138 | 1 156 | 1 137 | 1 243 | 3.9 | 0.6 | 11.9 | 4 |
| Maïs | 46 894 | 9 575 | 104 554 | 32 961 | 50 132 | 12 858 | 125 992 | 41 445 | 1 069 | 1 343 | 1 205 | 1 257 | 9 | 6.8 | 8.5 | 7 |
| Riz | 84 306 | 19 253 | 43 964 | 26 001 | 113 664 | 26 116 | 53 595 | 32 661 | 1 348 | 1 356 | 1 219 | 1 256 | 15.8 | 10.2 | 4.5 | 3.3 |
| Manioc | 5 404 | 2 716 | 11 635 | 10 635 | 7 220 | 3 378 | 11 502 | 7 496 | 1 336 | 1 244 | 989 | 705 | 22.4 | 10.8 | 11.3 | 1.5 |
| Arachide | 10 587 | 139 156 | 710 114 | 51 224 | 11 989 | 146 103 | 724 396 | 62 475 | 1 132 | 1 050 | 1 020 | 1 220 | 5.3 | 9 | 8.5 | 9 |
| Niébé | 13 520 | 48 797 | 85 816 | 7 065 | 13 156 | 50 812 | 75 418 | 5 879 | 973 | 1 041 | 879 | 832 | 1.6 | 3.7 | 1.8 | 1.2 |
| Noix de cajou | 81 | 5 541 | 699 | 13 777 | 53 | 5 424 | 1 028 | 16 765 | 655 | 979 | 1 470 | 1 217 | 2.2 | 5.5 | 2.1 | 1.5 |
| Coton | 12 | 4 977 | 565 | 13 108 | 6 | 5 097 | 832 | 16 025 | 541 | 1 024 | 1 473 | 1 223 | 6.7 | 9.8 | 19.5 | 12.5 |
| Pastèque | 2 | 15 507 | 30 | 471 | 3 | 17 947 | 18 | 443 | 1 522 | 1 157 | 586 | 941 | 14 | 13.8 | 26.2 | 0.7 |
| Oignon | 9 | 14 139 | 171 | 1 495 | 29 | 62 197 | 591 | 5 814 | 3 273 | 4 399 | 3 458 | 3 889 | 375 | 42.4 | 26.9 | 12.7 |
| Aubergine | 61 | 10 679 | 1 639 | 2 914 | 80 | 14 192 | 1 949 | 2 996 | 1 306 | 1 329 | 1 189 | 1 028 | 78.5 | 11.6 | 15.3 | 5 |

Source: résultats du modèle

**Tableau 13**: Revenu moyen agricole des exploitations pour la baseline selon les régions du Sénégal (FCFA)

| | Sénégal | Dakar | Thiès | Diourbel | Fatick_Kao | Kaffrine | St-Louis | Louga | Matam | Tamba_Ked | Casamance |
|---|---|---|---|---|---|---|---|---|---|---|---|
| Revenu moyen des exploitations | 1 386 438 | 1 837 832 | 1 786 112 | 1 043 783 | 970 203 | 2 151 235 | 2 819 904 | 1 236 324 | 404 379 | 1 050 591 | 1 427 194 |

Source: résultats du modèle

**Tableau 14**: Revenu moyen agricole des exploitations pour la baseline selon la taille économique des exploitations et leur spécialisation (FCFA)

|  | Petite | Moyenne | Grande | Vivrier | Rente | Ceréale/legumineuse | Mixte |
|---|---|---|---|---|---|---|---|
| Revenu moyen des exploitations | 411 893 | 1 238 550 | 3 316 865 | 757 875 | 2 325 430 | 1 287 355 | 2 135 574 |

Source: résultats du modèle

## 5.2 Effet sur l'utilisation des sols

Les effets des trois scénarios sur les assolements au niveau national sont présentés dans les Figure 3, Figure 4, et Figure 5. Avec des variations au niveau national comprises entre −1% (baisse de la surface de sésame dans le cas du scenario Univ) et +0.41% (surface de bissap pour le scenario Abol), les effets sur les assolements semblent relativement limités. Par exemple, dans le cas d'une suppression totale du programme de subvention actuel (Abol), les cultures à forte taux d'utilisation d'engrais comme le riz, la pastèque, la tomate et les autres cultures vivrières horticoles[2] sont partiellement substituées par des cultures à faible taux comme le sésame, le niébé et le bissap. Ces résultats s'expliquent par le fait que les cultures à forte intensité d'engrais deviennent moins compétitives avec la suppression des subventions et, par conséquent, perdent une partie de leur superficie au profit de cultures moins dépendantes des engrais. Au contraire, une extension du programme à l'ensemble des agriculteurs favoriserait très légèrement les surfaces du mais, du riz, de la tomate, et du coton (entre +0.1 et +0.2%) au détriment du sésame et du bissap qui sont cultivés sans pratiquement recours aux engrais (−0.95% et −0.65% respectivement). Dans le cas du scénario qui ciblerait les exploitations de moins de 5 ha, ce sont essentiellement le sésame et les autres cultures vivrières qui perdraient des hectares, maïs légèrement moins que dans le second scénario (−0.75% et −0.70% respectivement) au profit du maïs et du riz (+0.21% et +0.14%). Les deux derniers scenarios (Univ et Cibl) semblent favoriser la culture du maïs en raison de sa forte dépendance aux engrais mais aussi parce que c'est une culture pratiquée par la plupart des exploitants. En effet, environ 400 ha de maïs de plus sont à enregistrer dans ces deux scenarios.

Le riz semble réagir le mieux à la suppression des subventions, en termes de changement de superficie, bien qu'il ait un taux d'utilisation d'engrais assez élevé dans la baseline. Cela signifie que le riz reste très compétitif même avec l'augmentation du prix des engrais (c.à.d. suppression des subventions). A cela s'ajoute le fait que la substitution du riz par d'autres cultures pour certaines exploitations reste coûteuse pour ne pas dire techniquement impossible, du moins, à court et moyen termes.

**Figure 3**: Impact du scénario Abol sur les assolements (% de variation par rapport à la baseline)

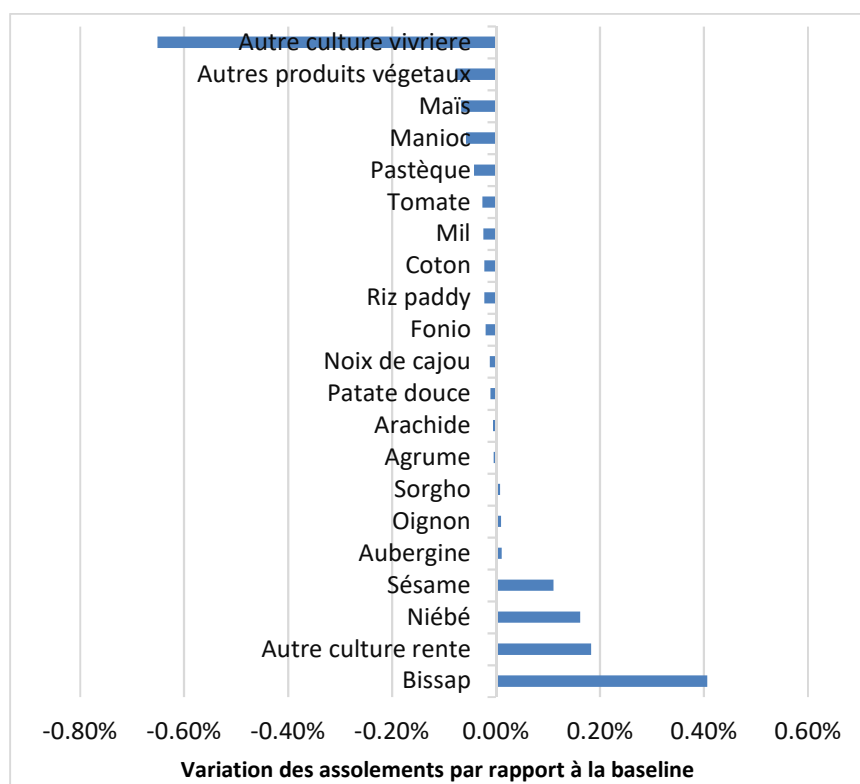

---

[2] Les activités "Autres cultures vivrières", "Autres produits végétaux", et "Autres culture de rente" sont incluses dans le questionnaire sans que davantage d'informations ne soient mentionnées sur la liste exacte des cultures formant ces activités. Nous avons fait le choix de conserver ces activités dans le modèle dans la mesure où l'objectif est d'intégrer l'ensemble des activités végétales observés dans la baseyear.

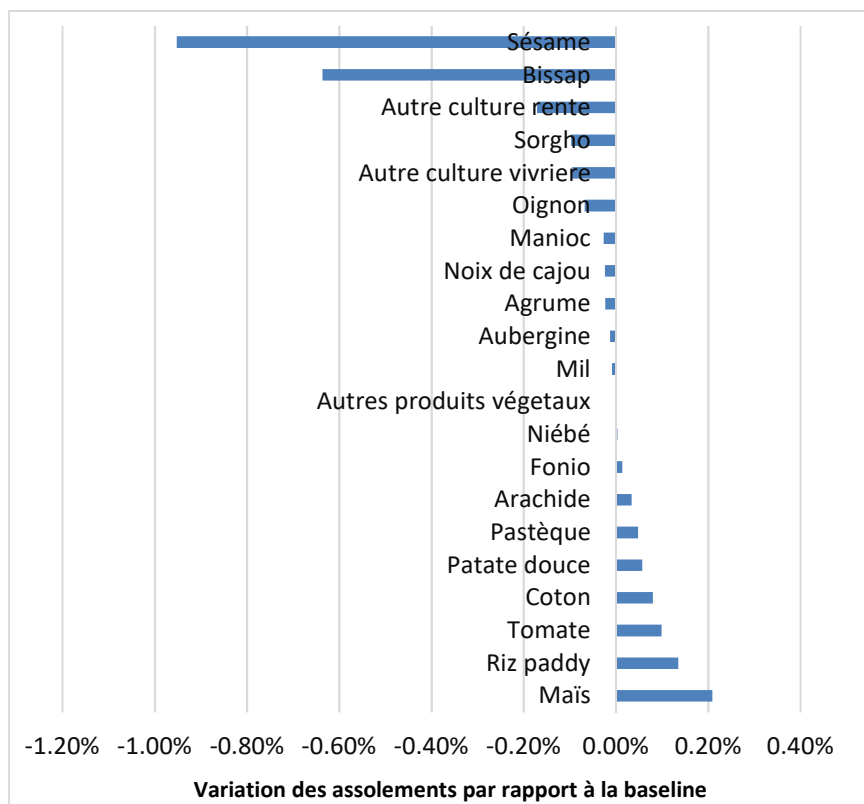

**Figure 4**: Impact du scénario Univ sur les assolements (% de variation par rapport à la baseline)

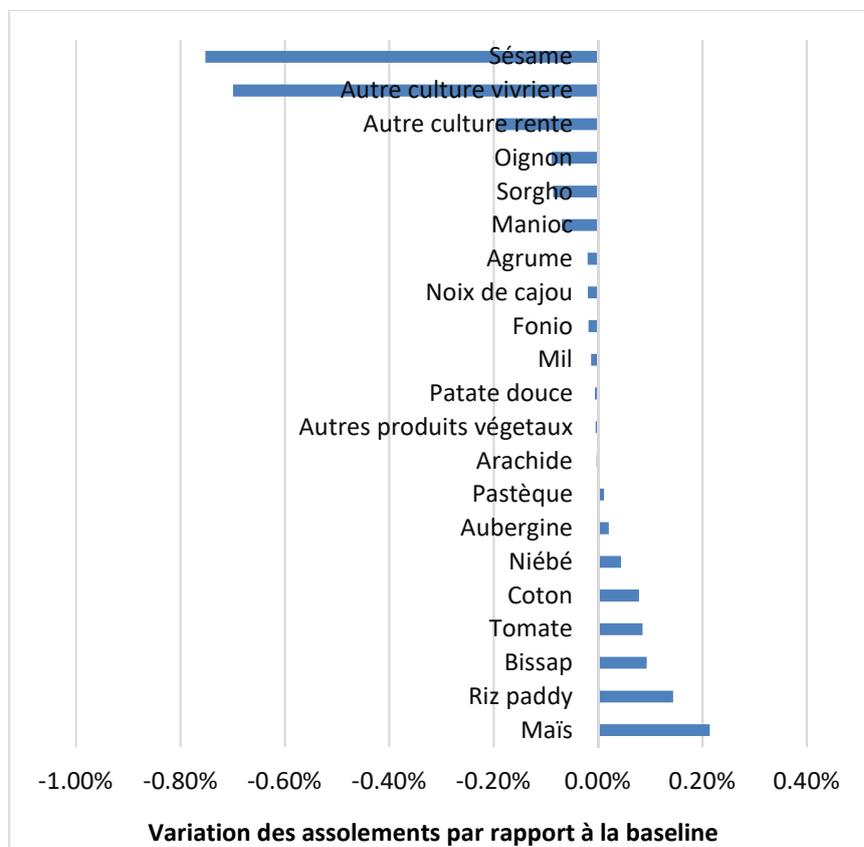

**Figure 5**: Impact du scénario Cibl sur les assolements (% de variation par rapport à la baseline)

## 5.3 Effet sur la production

### 5.3.1 Effets sur l'utilisation des engrais

Bien qu'un peu plus importants que dans le cas des assolements, les effets des scénarios sur les applications d'engrais restent très modérés au niveau agrégé, à quelques exceptions près. Ceci n'est pas le cas au niveau de l'exploitation individuel, où ces effets peuvent atteindre pour certaines d'entre elles des pourcentages à 2 chiffres.

A l'échelle du pays, une suppression des subventions aux engrais (scénario Abol) induirait une baisse du recours à l'engrais de seulement -0.14%. Cela parait relativement faible étant donné que le scénario Abol aboutit à supprimer le programme de subvention aux 30% des exploitants qui en bénéficient dans la baseline. Une explication est que les quantités subventionnées par ménage étaient relativement faibles, et donc leur suppression a un effet au final assez limité en termes de réduction d'engrais appliqué par les agriculteurs. Par ailleurs, on remarque que la plupart des exploitations continuent à utiliser les engrais même avec la suppression des subventions mais avec une moindre intensité. Une désagrégation des résultats à l'échelle régionale, montre que l'abolition du programme à des effets relatifs similaires entre les régions, allant de -0.03% (Fatick_kaolack) à -0.7% dans le cas de St-Louis dont le taux de bénéficiaire du programme actuel est un des plus élevés.

Une extension du programme à l'ensemble des ménages (scénario Univ) induit une hausse de l'utilisation des engrais de +0.65%. Le troisième scénario qui consiste à étendre le programme aux seules exploitations de moins de 5 ha (scénario Cibl) conduit quant à lui à une hausse de l'utilisation de +0.6% (Figure 6). Au niveau régional, ces deux scénarios ont un impact très fort sur la région de Matam avec une hausse des engrais par ha de +43% et +42% pour les scénarios Univ et Cibl, respectivement. Deux raisons expliquent ces taux de variation importants. Tout d'abord, le taux de bénéficiaires dans cette région dans le système actuel (baseline) est le plus faible (7.3%) de sorte qu'une extension du programme à l'ensemble des producteurs (Univ) ou aux exploitants ayant moins de 5 ha (Cibl) augmente drastiquement le nombre de bénéficiaires potentiels et, donc, l'effet sur l'utilisation des engrais. La seconde raison est le très faible usage des engrais dans cette région dans la baseline, ce qui induit des taux de variation assez élevé alors que la hausse en valeur absolue reste modeste. Ensuite, ce sont les régions de Dakar (+2.3%), Kaffrine (+1%) et de St Louis (+1%) qui sont les plus impactées par les scénarios Univ et Cibl.

Figure **7** présente les effets sur l'usage des engrais selon les cultures. Celles qui sont les plus impactées par la suppression du programme de subvention (Abol) sont le sorgho, le maïs, le gombo et les autres cultures vivrières et de rente (autour de -0.5% pour chaque culture). A l'inverse, les cultures pour lesquelles l'emploi d'engrais augmente le plus dans les scénarios Univ et Cibl sont le maïs, le riz, le niébé, le gombo, et les autres cultures de rente avec des hausses comprises entre +2% et +4.5%.

Enfin, Figure 8 présente l'impact des scénarios selon la taille économique des exploitations et leur spécialisation. D'après cette Figure il apparait clairement que les petites exploitations et celles spécialisées dans les cultures vivrières enregistrent la plus grande variation en terme relatif dans les trois scénarios. Dans le cas d'une suppression du programme de subvention de l'engrais, l'utilisation de l'engrais baisse de -0.7%, -0.1%, et -0.05% pour les petites, moyennes, et grandes exploitations, respectivement. Ce résultat est attendu et montre bien que les subventions facilitent l'accès aux engrais, notamment pour les petites exploitations.

Une généralisation du programme à l'ensemble des producteurs (Univ) conduirait à une hausse de +3.7%, +0.1%, et +0.1% pour les petites, moyennes, et grandes exploitations, respectivement. La réponse relativement forte chez les petites exploitations dans ce scénario est expliquée par l'augmentation du nombre de bénéficiaires dans ce groupe mais aussi par leurs faibles taux d'utilisation d'engrais dans la baseline.

Enfin, le ciblage plus restrictif (Cibl) conduirait à la même hausse pour les petites exploitations (+3.7%) que dans le cas du scénario Univ, à la même utilisation d'engrais pour les moyennes exploitations, et enfin à une très légère baisse des engrais pour les grandes exploitations (-0.1%). Dans ce dernier cas, c'est parce que la très grande majorité des exploitations ayant une grande taille économique (telle que définit dans cette étude) ont aussi une superficie supérieure à 5 ha et ne peuvent donc pas bénéficier du programme.

L'analyse des résultats selon la spécialisation des exploitations révèle que les principales perdantes d'un arrêt du programme (Abol) seraient les exploitations vivrières avec une baisse de -0.7% due notamment à la réduction de leurs superficies en maïs, riz et niébé. Dans le cas d'une extension du programme (Univ et Cibl), l'application d'engrais augmenterait dans les simulations de +3.1%. Les exploitations spécialisées en cultures de rentes ou bien les exploitations dites mixtes bénéficieraient, respectivement, de +0.7% et +0.5% d'une extension du programme à l'ensemble des ménages agricoles (Univ), mais de seulement 0.65% et 0.25% dans le cas du scénario Cibl.

**Figure 6**: Effets des scenarios simulés sur l'utilisation des engrais (kg/ha) à l'échelle nationale et régionale (% de variation par rapport à la baseline)

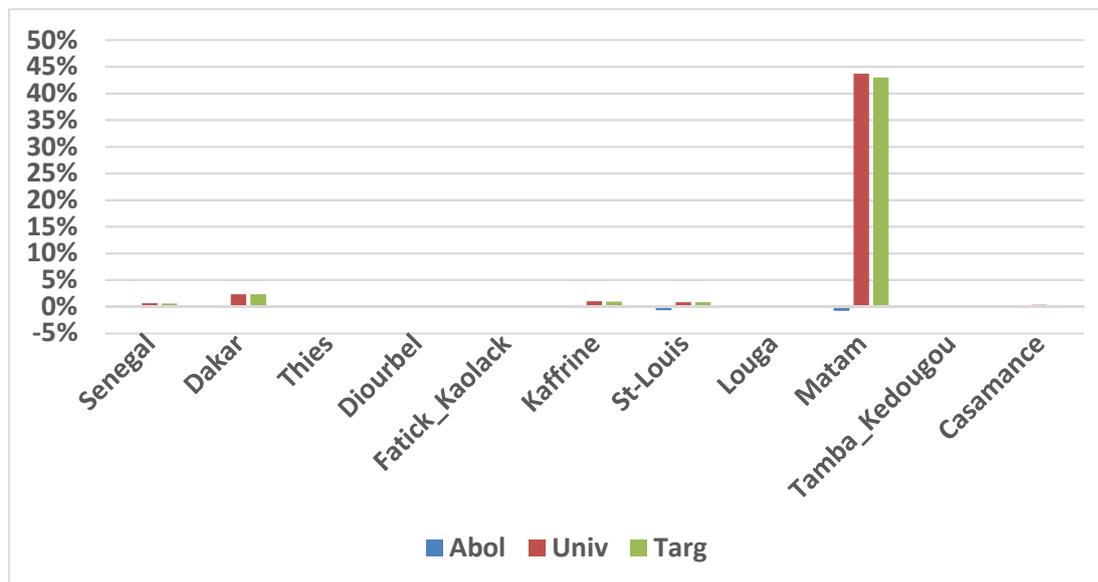

**Figure 7**: Effets des scenarios simulés sur l'utilisation des engrais (kg/ha) selon les cultures (% de variation par rapport à la baseline)

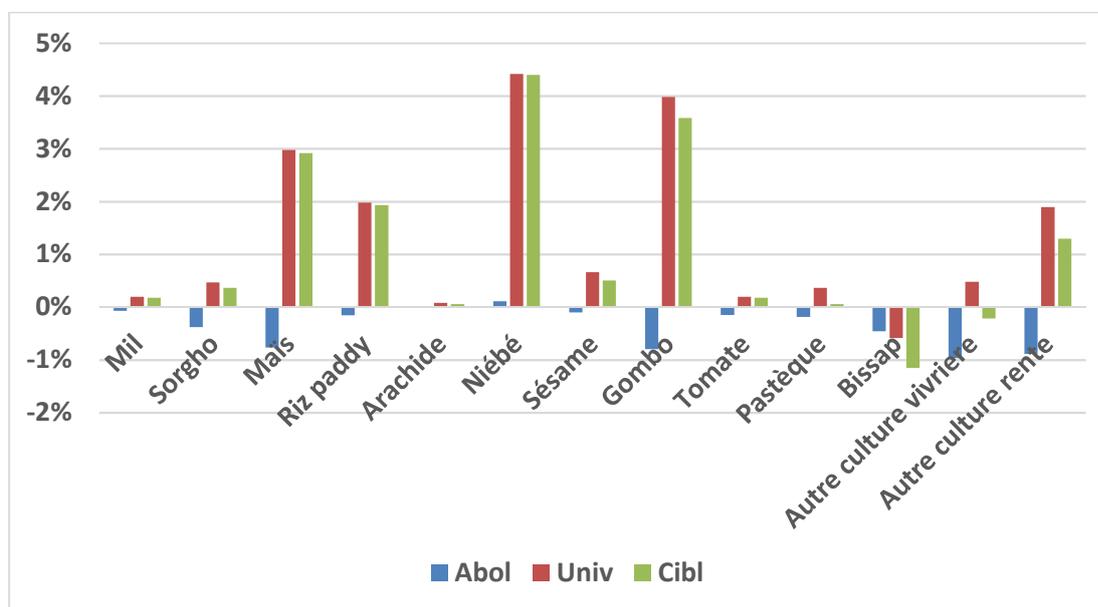

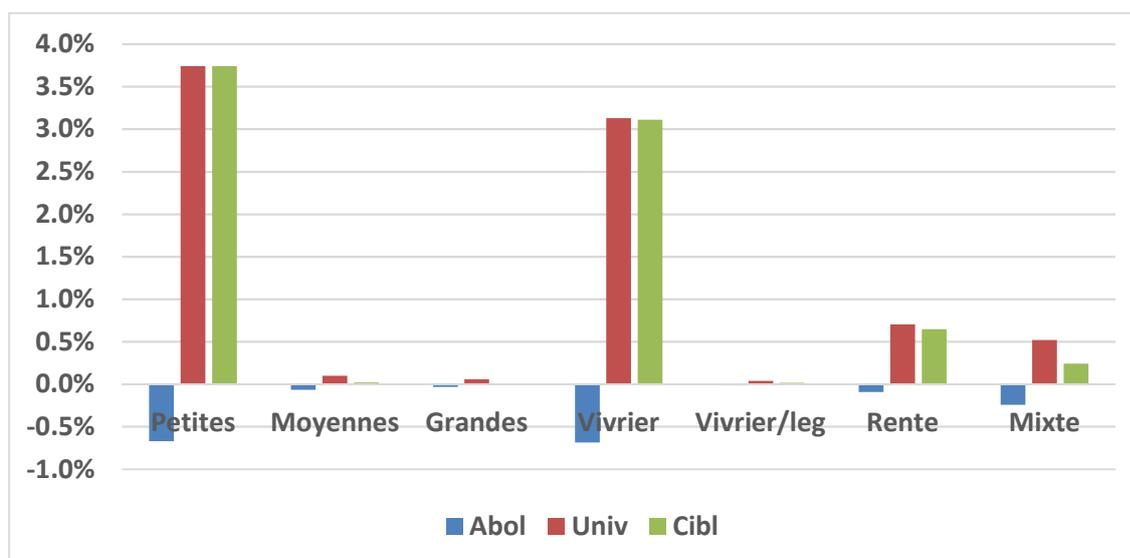

**Figure 8**: Effets des scenarios simulés sur l'utilisation des engrais (kg/ha) selon la taille économique et la spécialisation des exploitations (% de variation par rapport à la baseline)

### 5.3.2 Effets sur les volumes produits

La Figure 9 présente les effets les effets des trois scénarios étudiés sur les volumes de production. Ces effets seraient plutôt attribuables aux changements des assolements que des rendements. En effet, on observe pratiquement les mêmes tendances que pour les changements des assolements mais avec quelques différences en termes de magnitude pour certaines cultures. La seule exception est le riz dont les changements de volumes de production sont plutôt liés aux rendements et non aux assolements.

Comme pour l'assolement, les scénarios simulés ont très peu d'effets sur les céréales, notamment le mil et le sorgho. Seulement le maïs et le riz qui seraient légèrement impactés du fait de leurs taux d'utilisation d'engrais relativement élevés. Dans le cas des scénarios Univ et Cibl, ces deux dernières cultures voient leur production augmenter de +0.13%, ce qui reste malgré tout marginale. Les autres cultures qui bénéficient le plus du ciblage universel ou du ciblage sur les petites exploitations sont la tomate, les autres cultures de rente, et les autres produits végétaux. Sous le seul scénario Univ, le coton et l'arachide voient leur production légèrement augmenter (+0.06%). Sous ces deux scénarios, la production de sésame baisserait légèrement, de −0.7% (Univ) et −0.55% (Cibl), tout comme celle de manioc avec une baisse de −0.05% et −0.1% sous les scénarios Univ et Cibl respectivement.

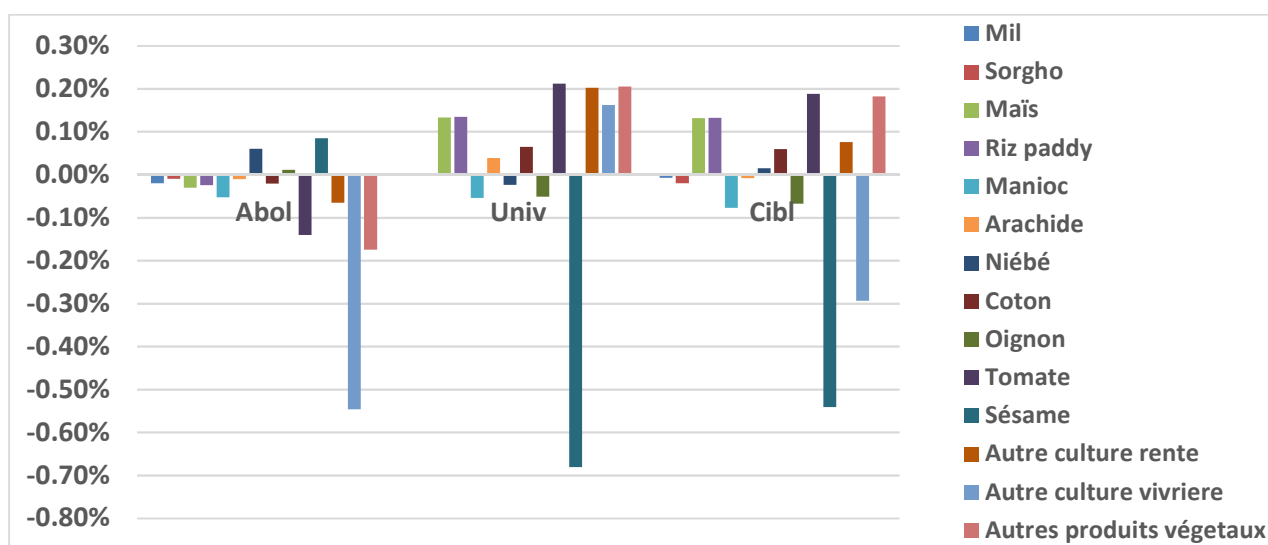

**Figure 9**: Effets des scenarios simulés sur les quantités produites (% de variation par rapport à la baseline)

## 5.4 Effets économiques

Les impacts des scénarios sur le revenu agricole à l'échelle du Sénégal et de ses régions, par type d'exploitation et par exploitation individuelle sont présentés dans les Figure 10, Figure 11, et Figure 12, respectivement.

La Figure 10 montre que les effets sur le revenu de trois scénarios simulés restent limités allant de -0.65% à +0.16 %, expliqués en grande partie par le faible taux d'utilisation des engrais dans les systèmes de production actuels. En raison de la faible part des coûts des engrais dans les coûts de production totaux, la suppression ou la réallocation des subventions aux engrais n'entraîne pas un impact aussi important sur les revenus agricoles. Cependant, un examen de plus près au niveau régional ou par type d'exploitation révèle des impacts plus prononcés et parfois de signes opposés.

D'après la Figure 10, on constate que les scénarios Abol et Cibl ont un effet négatif sur le revenu agricole de -0.65% et -0.28%, respectivement, alors que le scénario Univ, qui suppose une extension des subventions de l'engrais à l'ensemble des agriculteurs à un effet positif, bien que très modeste de +0.16%. Dit autrement, ce résultat révèle que le programme de subvention actuellement en place augmente le revenu agricole au niveau national de +0.65%, celui du programme de subvention ciblé de +0.37% et celui du programme de subvention universel de +0.81%.

Le résultat du scénario Cibl est étonnant, à première vue, dans la mesure où il a été montré plus en haut que mis à part la région de Kaffrine, il y a plus de ménages qui peuvent bénéficier du programme sous le scénario Cibl que dans la baseline. Néanmoins, la manière dont les subventions sont distribuées entre les exploitations dans ce scénario explique en grande partie ce résultat. En effet, les « grandes » exploitations dont la superficie est supérieure à 5 ha et qui sont à priori les plus compétitives sont exclues des subventions dans ce scénario et par conséquent voient leurs revenus baisser ce qui se traduit par une réduction du revenu global national.

Si on se concentre sur les résultats régionaux, on remarque que l'effet des subventions sur le revenu reste faible dans toutes les régions même s'il y a une petite disparité régionale. Sous le scénario Cibl, le revenu agricole moyen pour la région de Kaffrine baisse de -0.71% comme attendu puisqu'il y a moins de bénéficiaires que dans la baseline, mais il diminue aussi dans les régions de Fatick/Kaolack (-0.59%), Casamance (-0.29%), Thiès (-0.23%), et Tambakounda/Kedougou (-0.09%). Pour les autres régions, le revenu moyen des exploitations augmente, notamment à Matam ou la hausse est de +0.62%. On constate également que le revenu moyen augmente dans les régions à dominance des exploitations de petites tailles et diminue dans les régions à forte concentration d'exploitations de grandes tailles.

Par ailleurs, dans le scénario Abol, les régions où le revenu moyen baisse le plus sont celles ayant un taux de bénéficiaire le plus élevé comme Fatick/Kaolack (-0.92%), Kaffrine (-0.91%), Dakar (-0.74%), Diourbel (-0.64%), et St Louis (-0.62%). Enfin, les régions qui seraient le plus avantagé par le scénario Univ en termes de hausse de revenu sont Dakar (+0.94%), Matam (+0.72%) et enfin St Louis (+0.46%).

**Figure 10**: Effets des scenarios simulés sur le revenu agricole à l'échelle nationale et régionale

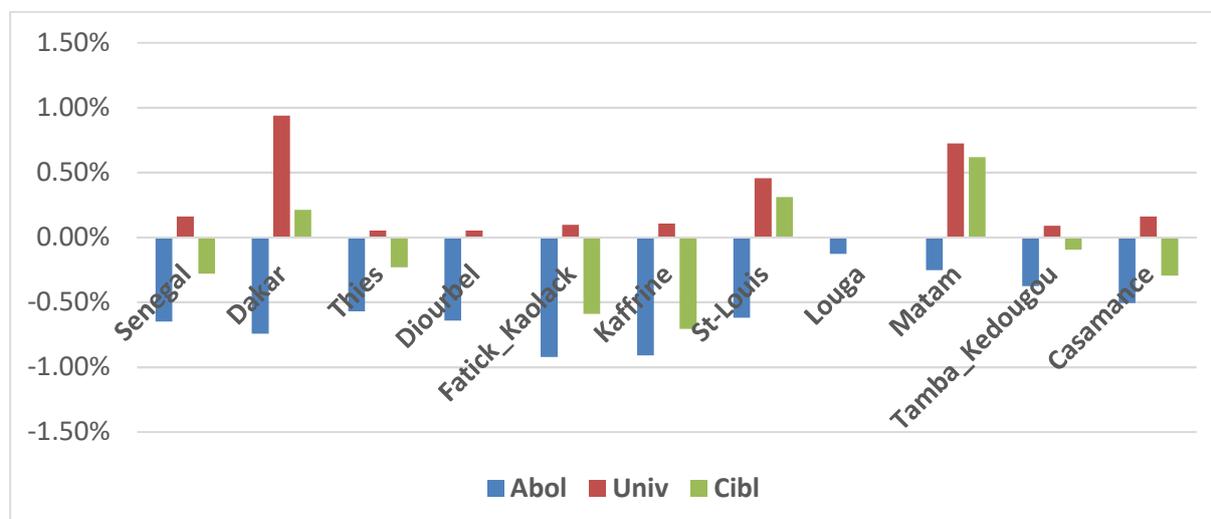

La Figure 11 présente les effets des scénarios sur le revenu agricole selon la taille économique des exploitations et leur spécialisation. En cas de suppression totale du programme de subvention (Abol), ce sont les plus petites exploitations qui perdraient le plus de revenu en terme relatif (-1.02%), mais aussi les exploitations spécialisées en cultures vivrières (-2.71%), suivi par les exploitations mixtes (-0.96%), puis les exploitations spécialisées en céréales/légumineuses. Ce sont ces mêmes types d'exploitations qui profiteraient le plus du scénario Univ et Cibl. Sous le scenario Cibl, les grandes exploitations perdraient en moyenne -0.43% de leur revenu, et les exploitations moyennes -0.16%.

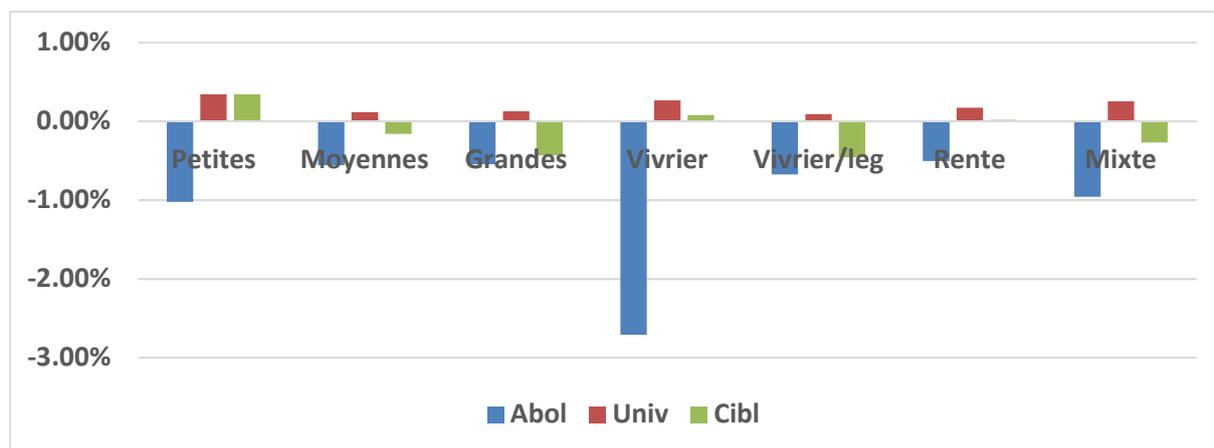

**Figure 11**: Effets des scenarios simulés sur le revenu agricole selon la taille économique des exploitations et leur spécialisation

La Figure 12 donne les écarts de revenus (en pourcentage), par exploitation individuelle, entre la baseline et le scénario Abol, entre le scénario Cibl et le scénario Abol et enfin entre le scénario Univ et le scénario Abol. Le choix du scénario Abol pour le calcul des écarts a pour but, d'une part, d'avoir uniquement des écarts nuls ou positifs et, d'autre part, de pouvoir comparer les effets distributifs des trois types de ciblage : programme de subvention actuel (baseline), programme de subvention universel (Univ) et programme de subvention ciblé (Cibl).

Ce qui apparait en premier lieu sur cette figure c'est que pour le scénario Univ, pour à peu près 55% des ménages agricoles, il n'y a pas de changement de revenu. Cette valeur monte à 68% pour la baseline et même à 74% sous le scénario Cibl. Si on déduit de ces pourcentages les ménages non-concernés (non-bénéficiaires) dans chaque scénario et qui sont 0% pour Univ, 38% pour Cibl et 69% pour la baseline, nous donne des taux de ménages ayant droit (concernés) mais non-affectés d'environ 55% pour Univ, 38% pour Cibl et 0% pour baseline (mais dans la baseline il n'est en réalité pas possible de mesurer le nombre d'ayant droit qui n'ont pas utilisés les subventions). Cela signifie qu'il y a un pourcentage assez élevé de ménages agricoles, notamment dans le scénario Univ, qui ne sont pas en mesure d'accéder à l'engrais subventionné malgré qu'ils aient droits à cause de la contrainte de liquidité.

On constate également que la majorité des ménages (90%) n'augmente leurs revenus que légèrement (moins de 5%) avec les programmes de subventions. Les ménages restant (10%) voient leurs revenus augmenter entre 5% et 40%, ce qui n'est pas négligeable. Ce sont en réalité ces 10% qui sont en mesure d'accéder aux engrais à un prix subventionné dont ils ont droit (c.à.d. jusqu'au quota). Pour les autres ménages, soit ils ne sont pas concernés (non-bénéficiaires), soit ils sont concernés mais leur contrainte budgétaire ne leur permet pas d'acheter des engrais à un prix subventionné, soit ils sont concernés mais leur contrainte budgétaire leur permet d'acheter qu'une petite quantité d'engrais subventionné (c.à.d. ne remplissent pas leur quota).

Cette Figure 12 montre également que :(i) le programme de subvention universel (Univ) a un meilleur effet distributif en termes de variation de revenu, ce qui est attendu vue son caractère universel. (ii) le programme de subvention ciblé améliore la distribution par rapport au programme actuel (Baseline) pour environ 15% de la population mais pénalise environ 17% de la population et (iii) les effets distributifs des scénarios Cibl et Univ sont très similaires pour les 5% de la population qui bénéficient le plus des subventions.

**Figure 12**: Distribution de la variation de revenu des ménages agricoles dans la baseline, le scénario Univ et le scenario Cibl, par rapport au scénario Abol (%)[1]

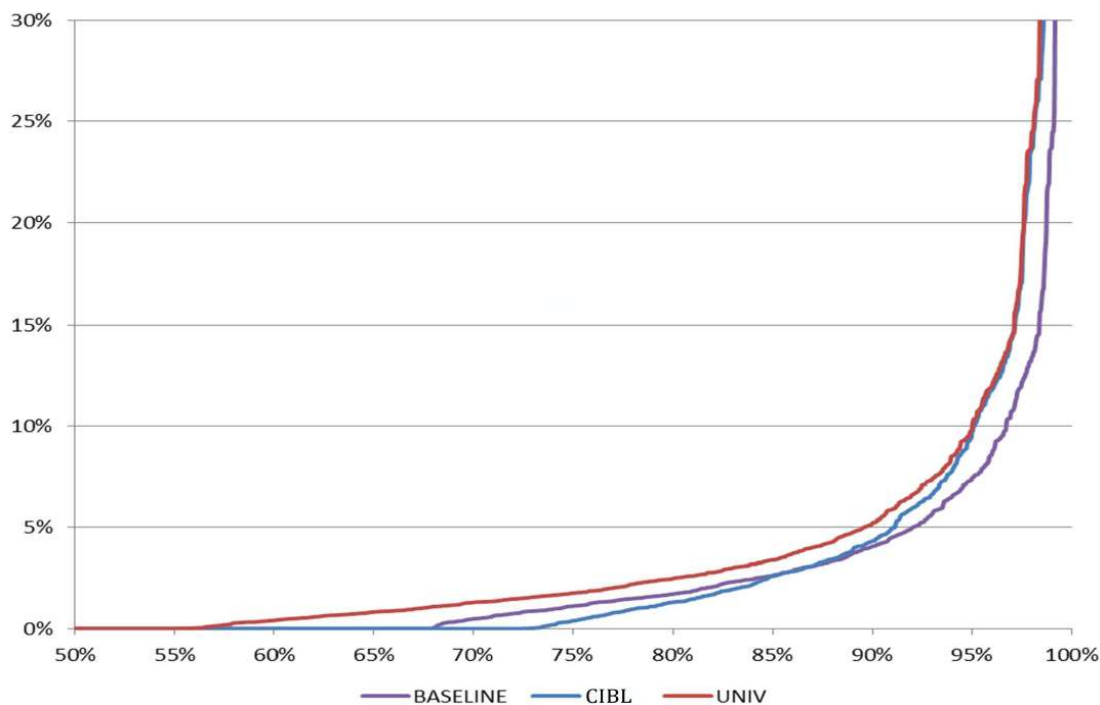

[1] Axe vertical: écart de revenu par rapport au scénario Abol (en pourcentage) ; axe horizontal : pourcentage cumulé de ménages agricoles.

Connaitre les effets des programmes (scénarios) simulés sur le revenu des agriculteurs est important mais n'est pas suffisant pour décider lequel il faut choisir. Il faut également estimer leurs coûts budgétaires. En effet, pour un gouvernement connaitre les coûts d'un programme est cruciale non seulement pour pouvoir estimer son degré d'efficacité mais aussi pour réfléchir sur les potentielles sources de financement. Du point de vue de l'analyse coûts-bénéfices, le programme le plus efficace est celui qui présente un ratio coûts-bénéfices le plus élevé, c'est-à-dire celui qui permet générer le bénéfice le plus élevé au moindre coût. Un ratio coûts-bénéfices inférieur à 1 indique que les coûts du programme dépassent le total de ses retombées nettes, ce qui implique son échec.

Dans cette étude en raison d'un manque de donnés, nous avons réduit les bénéfices aux gains de revenus des ménages agricoles et les coûts aux montants des subventions aux engrais accordés par l´Etat aux ménages agricoles à travers une réduction des prix. Cela signifie, par exemple, que les coûts administratifs de mise en œuvre des programmes ne sont pas pris en compte. Dans une telle situation, ce ratio peut être interprété comme le gain en termes de revenu généré par 1 FCFA apporté par l'Etat par le biais d'une subvention aux l'engrais.

La Figure 13 présente les coûts et les bénéfices des trois programmes (scenarios) : le programme de subvention actuellement en place (baseline), le programme de subvention universel (Univ) et le programme de subvention ciblé (Cibl). Le scénario Abol, qui correspond à la situation sans programme de subvention (0 coût et 0 bénéfice) n'est donc pas représenté sur la figure. Dans le cas de la baseline, le coût budgétaire tel que calculé dans le modèle est de 3.88 milliards de FCFA pour un bénéfice total de 4.08 milliards de FCFA. Sous le scénario Univ, le coût est de 4.64 milliards de FCFA pour un bénéfice estimé à 5.17 milliards de FCFA. Enfin, sous le scénario Cibl, le coût budgétaire est de 1.99 milliards pour un bénéfice estimé à 2.39 milliards de FCFA. Avec un ratio coûts-bénéfices de 1.2, le scénario Cibl semble être le plus efficient, suivi par le scenario Univ (ratio de 1.11), et enfin la baseline (ratio de 1.05). Cela signifie que, même si la différence entre les trois programmes n'est pas assez significative, cibler les petites exploitations, qui sont aussi généralement celles qui ont les plus fortes productivités marginales de l'engrais, s'avère être l'option la plus efficace.

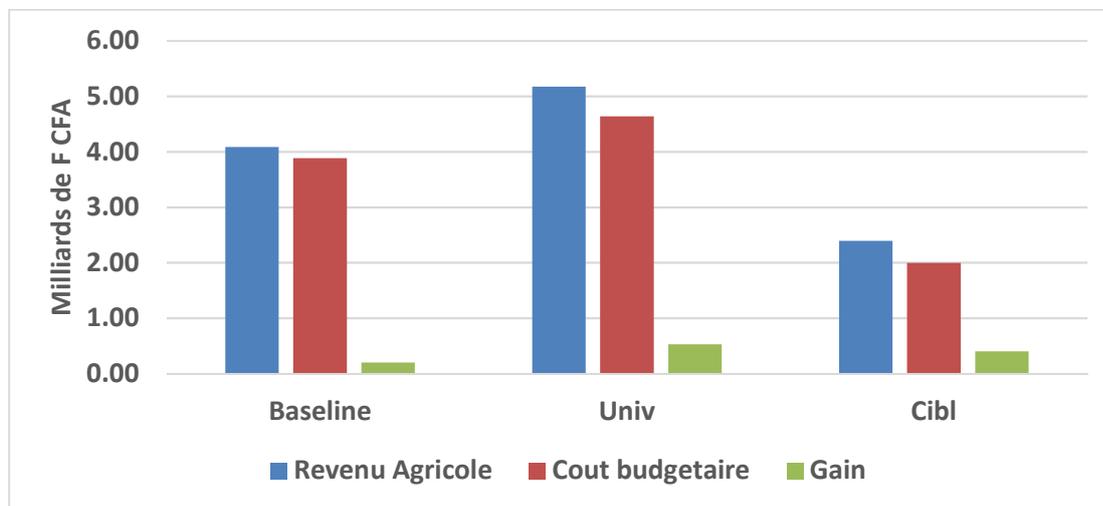

**Figure 13**: Comparaison des bénéfices et des coûts engendrés sous chacun des scénarios simulés

# 6 Conclusions

Le mécanisme de ciblage des bénéficiaires du programme de subvention des intrants au Sénégal, tel qu'il existe aujourd'hui, est de plus en plus critiqué par de nombreux acteurs des filières agricoles. Avec des quantités d'intrants subventionnés relativement limitées à l'échelle du pays, tous les agriculteurs ne peuvent pas bénéficier de ces intrants de sorte qu'un ciblage se réalise de facto au travers du filtre des commissions locales. Ces commissions locales regroupent différents acteurs impliqués dans le secteur agricole à l'échelle d'une (ou plusieurs) commune(s) donnée(s). Ces acteurs sont généralement des responsables politiques locaux, des personnes impliquées dans les organisations villageoises, dans la collecte de produits agricoles, etc. Etant donné que ces commissions locales ne procèdent pas nécessairement à un ciblage systématisé et codifié, les critères d'attribution des intrants subventionnés aux agriculteurs sont donc peu connus et relativement opaques. Le gouvernement sénégalais ainsi que les bailleurs de fond participant au financement de ce programme souhaitent réformer le programme afin d'aboutir à un ciblage plus transparent des bénéficiaires.

Ce rapport présente les résultats d'une évaluation d'impact de ce programme et notamment une comparaison avec deux programmes alternatifs basés sur différent mode de ciblages. Trois scénarios ont été ainsi simulés : (i) une suppression totale du programme actuel ce qui nous permet d'apprécier ses effets, (ii) une extension du programme actuel à l'ensemble des ménages agricoles sans aucune distinction mais avec une réduction de moitié du quota d'engrais subventionné par ménage (ciblage de type universel), et enfin (iii) un ciblage plus plausible qui donnerait accès aux seules exploitations agricoles de moins de 5 ha.

Cette évaluation a été réalisée par le biais du modèle de ménage agricole FSSIM-Dev basé sur la programmation mathématique positive. Les données issues de la seconde Enquête de Suivi de la Pauvreté au Sénégal (ESPS2) réalisée en 2011 ont été utilisées dans cette évaluation d'impact. Ces données présentent l'avantage de couvrir l'ensemble du pays ce qui a permis de générer des résultats représentatifs aussi bien niveau régional que national. En plus, les simulations sont réalisées pour chaque ménage agricole présent dans la base de données, ce qui a permis d'avoir des résultats à l'échelle individuelle puis de les agréger à l'échelle régionale puis nationale en utilisant un facteur de pondération adéquat.

Les principaux résultats de cette évaluation sont les suivants :

i) Que ce soit au niveau national ou régional, les différents modes de ciblage du programme de subvention simulés ici affectent peu les assolements, le taux d'utilisation d'engrais, les volumes produits, ainsi que le revenu agricole (allant de -0.65% à +0.16 %) ;

ii) Au niveau individuel, les effets peuvent être plus prononcés. Pour certains ménages agricoles, l'effet sur le revenu pourrait aller jusqu'à +30% ;

iii) Plusieurs ménages (environ 45% dans le cas d'une universalisation du programme) sont contraints par la trésorerie de continuer à produire sans engrais, malgré les subventions ;

iv) Au niveau régional, les effets les plus importants sont observés dans les régions de Matam, St Louis et Dakar en raison du grand nombre de bénéficiaires. La région qui serait la plus impactée par une extension du programme de subvention des engrais (scénarios Univ) est Matam (+40%). C'est aussi la région dont le taux de bénéficiaire actuel (baseline) est le plus faible. ;

v) Les cultures qui bénéficieraient le plus du programme de subvention des engrais seraient principalement le maïs et le riz (+3%), ainsi que le niébé (+4%) ;

vi) Les petites exploitations et celles spécialisées en cultures vivrières seraient les plus dépendantes des subventions. Par exemple, dans le cas d'une suppression du programme, les petites exploitations et les exploitations vivrières perdraient le plus de revenu ;

vii) Selon l'analyse coûts-bénéfices, un programme qui cible les exploitations ayant moins de 5 ha semble le plus efficace (scénario Cibl). Son ratio bénéfice (mesuré en termes de gain de revenu) / coûts (mesurés en termes de subventions allouées) est de 1.2, soit le plus élevé.

Bien que les résultats de simulation présentés dans ce rapport soient le fruit d'un intense travail visant à modéliser le plus finement possible le fonctionnement des ménages agricoles au Sénégal, une grande prudence s'impose toutefois dans l'interprétation de ces résultats compte tenu des hypothèses et limitations du modèle mais aussi de la qualité des données utilisées. Parmi les limitations de ce modèle, on mentionnera en premier lieu l´absence d'effet marché. En effet, le modèle fonctionne avec des prix exogènes, par conséquent les effets des scénarios simulés sur les prix des produits et des facteurs de production, tout particulièrement l'engrais, ne sont pas pris en compte. Cependant, étant donné l'effet très limité de ces scenarios sur la production et sur la demande d'engrais, cette limitation reste secondaire. Plus important, certains éléments clés comme l'aversion au risque, les interactions entre exploitations pour certains marchés (foncier, travail, etc.), ou encore les défaillances du système de distribution des subventions n'ont pas été considérées dans cette évaluation. Ces éléments, notamment le dernier, risquent de biaiser partiellement les résultats ou entrainer une surestimation des impacts pour certains ménages et une sous-estimation pour d'autres. Enfin, le jeu de données utilisé, bien que couvrant l'ensemble du pays, présente plusieurs limites. Tout d'abord, il ne contient pas les grandes exploitations commerciales tournées vers l'exportation et qui pourraient aussi être affectées par une réforme du système de distribution des intrants subventionnés. Ensuite, les informations sur les types d'engrais utilisés par les agriculteurs pour chaque culture n'étaient pas fournies. Cela a empêché de modéliser explicitement la relation entre l'azote et les rendements via une fonction de production comme cela a été réalisé dans d'autres études utilisant le modèle FSSIM-Dev (Louhichi et al. 2019, Ricome et al. 2020). Ce manque de lien explicite entre les quantités d'engrais utilisés et les rendements explique en partie la rigidité du modèle et les effets assez limités des scénarios sur les volumes produits, et donc sur les revenus.

Malgré ces limitations, les résultats présentés dans ce rapport peuvent aider les décideurs politiques à mieux comprendre comment les ménages à divers endroits du pays peuvent être affectés par les programmes simulés et servir comme support pour la conception de futurs programmes. Ces résultats confirment également l'intérêt d'étendre l'analyse à d'autres intrants, tels que le matériel agricole ou les produits phytosanitaires qui bénéficient également de subventions, ou bien de tester d'autres types de ciblage, comme le ciblage régional (cibler par exemple les régions qui utilisent peu d'engrais) ou le ciblage sectoriel (cibler les secteurs/produits stratégiques).

Finalement, ce rapport met en exergue les potentialités de cette approche par la modélisation. Ce type d'approche permet, en effet, de produire des informations utiles pour les décideurs politiques intéressés par l'analyse d'impact des politiques agricoles à différents niveaux d'analyse, depuis l'échelle individuelle (les ménages agricoles) à l'échelle nationale, en passant par l'échelle régionale et le ménage type.

## **Références**

Banque Mondiale (2020). Indicateurs du développement dans le Monde. Site internet de la Banque Mondiale, Washington, Etats-Unis, Données extraites en Janvier 2020.

Boulanger, P., H. Dudu, E. Ferrari, A. Mainar, F. Angelucci, R. Baborska and T. Mailland (2018). *Allocations budgétaires optimales et options de réformes pour le secteur agricole dans le Plan Sénégal Emergent 2019-2023*, JRC Science for Policy Report, Publications office of the European Union, Luxembourg.

FAO (2019). FAOSTAT. FAO, Rome, Italie, Données extraites en Janvier 2020 sur le site internet de la FAO.

Feed the Future (2018). *La subvention des engrais au Sénégal: Revue et perspectives*. Série Note d'information 03,, Dakar, Sénégal**:** 7pp.

Heckelei, T. (2002). *Calibration and Estimation of Programming Models for Agricultural Supply Analysis*. Habilitation Thesis, Agricultural Faculty, University of Bonn, Bonn, Germany.

Heckelei, T., R. C. Mittelhammer and T. Jansson (2008). A Bayesian alternative to generalized cross entropy solutions for underdetermined econometric models. Discussion Papers 56973. Bonn, Institute for Food and Resource Economics.

Henry de Frahan, B., J. Buysse, P. Polomé, B. Fernagut, O. Harmignie, L. Lauwers, G. Van Huylenbroeck and J. Van Meensel (2007). Positive Mathematical Programming for Agricultural and Environmental Policy Analysis: Review and Practice. Handbook of Operations Research in Natural Resources. A. Weintraub, C. Romero, T. Bjørndal and R. Epstein, Springer US**:** 129-154.

Howitt, R. E. (1995). *A Calibration Method for Agricultural Economic Production Models*. Journal of Agricultural Economics 46(2): 147-159.

IPAR (2015). *Subventions des intrants agricoles au Sénégal: Controverses et réalités*, Dakar, Sénégal**:** 44pp.

Janssen, S., K. Louhichi, A. Kanellopoulos, P. Zander, G. Flichman, H. Hengsdijk, E. Meuter, E. Andersen, H. Belhouchette, M. Blanco, N. Borkowski, T. Heckelei, M. Hecker, H. Li, A. Oude Lansink, G. Stokstad, P. Thorne, H. van Keulen and M. K. van Ittersum (2010). *A Generic Bio-Economic Farm Model for Environmental and Economic Assessment of Agricultural Systems*. Environmental Management 46(6): 862-877.

Louhichi, K., P. Ciaian, M. Espinosa, A. Perni and S. Gomez y Paloma (2018). *Economic impacts of CAP greening: application of an EU-wide individual farm model for CAP analysis (IFM-CAP)*. European Review of Agricultural Economics 45(2): 205-238.

Louhichi, K. and S. Gomez y Paloma (2014). *A farm household model for agri-food policy analysis in developing countries: Application to smallholder farmers in Sierra Leone*. Food Policy 45: 1-13.

Louhichi, K., A. Kanellopoulos, S. Janssen, G. Flichman, M. Blanco, H. Hengsdijk, T. Heckelei, P. Berentsen, A. O. Lansink and M. V. Ittersum (2010). *FSSIM, a bio-economic farm model for simulating the response of EU farming systems to agricultural and environmental policies*. Agricultural Systems 103(8): 585-597.

Louhichi, K., U. Temursho, L. Colen and S. Gomez y Paloma (2019). *Upscaling the productivity performance of the Agricultural Commercialization Cluster Initiative in Ethiopia*, JRC Science for Policy Report, Publications office of the European Union, Luxembourg.

Louhichi, K., P. Tillie, A. Ricome and S. Gomez y Paloma (2020). *Modelling farm-household livelihood in Developing Economies: Insights from three country case studies using LSMS-ISA data*, JRC Science for Policy Report, Publications office of the European Union, Luxembourg.

Mérel, P. and S. Bucaram (2010). *Exact calibration of programming models of agricultural supply against exogenous supply elasticities*. European Review of Agricultural Economics 37(3): 395-418.

Ministère de l'Agriculture et de l'Equipement Rural (2014). Programme d'Accélération de la Cadence de l'Agriculture Sénégalaise (PRACAS) - Volet agricole du Plan Sénégal Emergent**:** 112pp.

Paris, Q. and R. E. Howitt (1998). *An Analysis of Ill-Posed Production Problems Using Maximum Entropy*. American Journal of Agricultural Economics 80(1): 124-138.

Ricome, A., K. louhichi and S. Gomez y Paloma (2020). *Impacts of agricultural produce cess (tax) reform options in Tanzania: A micro-economic analysis using a farm-household model*, JRC Science for Policy Report, Publications office of the European Union, Luxembourg.

Rioux, J., A. Gamli, K. Amourou and G. Ndiaye (2011). *Analyse Globale de la Vulnérabilité de la Sécurité Alimentaire et de la Nutrition*, République du Sénégal, Dakar, Sénégal**:** 181pp.

Sall, M. (2015). Les exploitations agricoles familiales face aux risques agricoles et climatiques: stratégies développées et assurances agricoles. PhD Thesis, Université de Toulouse, 578pp.

Seck, A. (2017). *Fertiliser subsidy and agricultural productivity in Senegal*. The World Economy 40(9): 1989-2006.

Tillie, P., K. Louhichi and S. Gomez y Paloma (2018). *La culture attelée dans le bassin cotonnier en Côte d'ivoire: Modélisation et analyse des impacts d'un programme de relance de la culture attelée*, JRC Science for Policy Report, Publications office of the European Union, Luxembourg.

Tillie, P., K. louhichi and S. Gomez y Paloma (2019). *Impacts ex-ante de la Petite Irrigation au Niger*, JRC Science for Policy Report, Publications office of the European Union, Luxembourg.

van Ittersum, M. K., F. Ewert, T. Heckelei, J. Wery, J. Alkan Olsson, E. Andersen, I. Bezlepkina, F. Brouwer, M. Donatelli, G. Flichman, L. Olsson, A. E. Rizzoli, T. van der Wal, J. E. Wien and J. Wolf (2008). *Integrated assessment of agricultural systems – A component-based framework for the European Union (SEAMLESS)*. Agricultural Systems 96(1): 150-165.

**Liste des abréviations**

| | |
|---|---|
| ANSD | Agence Nationale de la Statistique et de la Démographie |
| ESPS2 | Enquête de Suivi de la Pauvreté au Sénégal – 2ème vague |
| FAO | Food and Agriculture Organisation |
| FSSIM-Dev | Farm System Simulator for Developing Countries |
| GOANA | Grande Offensive Agricole pour la Nourriture et l'Abondance |
| IDH | Indice de Développement Humain |
| ISRA | Institut Sénégalais de la Recherche Agricole |
| MAER | Ministère de l'Agriculture et de l'équipement Rural |
| PIB | Produit Intérieur Brut |
| PNUD | Programme des Nations Unies pour le Développement |
| PRACAS | Programme d'Accélération de la Cadence de l'Agriculture Sénégalaise |
| PSE | Plan Sénégal Emergent |
| TS4FNS | Technical Support for Food and Nutrition Security |
| USD | US dollar |

## Liste des figures

## Liste des tableaux

**GETTING IN TOUCH WITH THE EU**

**In person**

All over the European Union there are hundreds of Europe Direct information centres. You can find the address of the centre nearest you at: https://europa.eu/european-union/contact_en

**On the phone or by email**

Europe Direct is a service that answers your questions about the European Union. You can contact this service:

- by freephone: 00 800 6 7 8 9 10 11 (certain operators may charge for these calls),

- at the following standard number: +32 22999696, or

- by electronic mail via: https://europa.eu/european-union/contact_en

**FINDING INFORMATION ABOUT THE EU**

**Online**

Information about the European Union in all the official languages of the EU is available on the Europa website at: https://europa.eu/european-union/index_en

**EU publications**

You can download or order free and priced EU publications from EU Bookshop at: https://publications.europa.eu/en/publications. Multiple copies of free publications may be obtained by contacting Europe Direct or your local information centre (see https://europa.eu/european-union/contact_en).

## The European Commission's science and knowledge service
Joint Research Centre

### JRC Mission
As the science and knowledge service of the European Commission, the Joint Research Centre's mission is to support EU policies with independent evidence throughout the whole policy cycle.

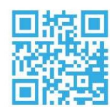 **EU Science Hub**
ec.europa.eu/jrc

@EU_ScienceHub

EU Science Hub – Joint Research Centre

EU Science, Research and Innovation

EU Science Hub

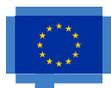
Publications Office
of the European Union